\documentstyle[aps,prabib,psfig]{revtex}
\onecolumn 
\tighten
\begin{document}

\title{
\begin{flushright}
\vspace{-1cm}
{\normalsize MC/TH 96/30\\
}
\vspace{1cm}
\end{flushright}
Coherent amplification of classical pion fields during the cooling of
droplets\\
of quark plasma}

\author{Abdellatif Abada and Michael C. Birse}

\address{ Theoretical Physics Group, Department of Physics and Astronomy,\\
University of Manchester, Manchester M13 9PL, England }

\date{December 2, 1996}

\maketitle

\begin{abstract}
In the framework of the linear sigma model, we study the time evolution of a
system of classical $\sigma$ and pion fields coupled to quarks. For this 
purpose we solve numerically the classical transport equation for relativistic 
quarks coupled to the nonlinear Klein-Gordon equations for the meson fields.
We examine evolution starting from variety of initial conditions
corresponding to spherical droplets of hot quark matter, which might mimic the
behaviour of a quark plasma produced in high-energy nucleus-nucleus
collisions. For large droplets we find a strong amplification of the pion
field that oscillates in time. This leads to a coherent production of pions
with a particular isospin and so would have similar observable effects to a
disoriented chiral condensate which various authors have suggested might be a
signal of the chiral phase transition. The mechanism for amplification of the
pion field found here does not rely on this phase transition and is better
thought of as a ``pion laser" which is driven by large oscillations of the
$\sigma$ field.
\end{abstract}

\section{Introduction}
\label{intro}

One of the main features of QCD, the underlying theory of strong interactions,
is the spontaneous breaking of its approximate SU(2)$\times$SU(2) chiral
symmetry. Spontaneous breaking of this approximate symmetry explains the
very small pion masses; in the limit of exact chiral symmetry these particles
would be massless Goldstone bosons. Another manifestation of this is the
presence of a non-vanishing quark condensate in the vacuum. It is believed
that at very high temperature a quark-gluon plasma is formed in which
chiral symmetry is restored, and much effort is being made to explore such a
phase transition by means of high-energy hadron or heavy-ion collisions.

To describe aspects of QCD related to this symmetry it is convenient to
introduce a chiral four-vector of fields $(\sigma, \mbox{\boldmath $\pi$})$,
where $\sigma$ represents the quark condensate and the three $\pi_i$ are the
pion fields. In the physical vacuum, $(\sigma, \mbox{\boldmath $\pi$})$
points in the $\sigma$ direction. If the chiral symmetry were exact then there
would be a ``chiral circle" of states degenerate with this vacuum states. In
practice the symmetry is explicitly broken by the current quark masses and so
there is a unique vacuum.

Because of this circle of nearly degenerate field configurations, as the
chirally restored plasma cools and returns to the normal phase the system
could form regions in which the chiral fields are misaligned, that is chirally
rotated from their usual orientation along the $\sigma$ direction. There has
been much recent interest in this phenomenon, which is known as a disoriented
chiral condensate (DCC). If such a state were formed, it would lead to
anomalously large event-by-event fluctuations in the ratio of charged to
neutral pions. Since the emergence of this idea\cite{Ans,Bjo,RW}, a lot of
theoretical work has gone into developing suitable methods for modelling
the phenomenon and in exploring how it could be used as a signal of a phase
transition in high-energy nucleus-nucleus or hadron collisions; reviews on the
subject can be found in Refs.~\cite{Raj,ARS}.

A region of DCC can also be thought of as a coherent state of low-momentum
pions. In order to produce such a state the hot plasma must evolve far from
equilibrium and in particular it must reach an unstable configuration in which
long-wavelength pion modes grow exponentially\cite{RW}. However it is not 
clear that the chiral phase transition is the only mechanism that could produce
such a coherent pion excitation. Moreover if there were others ways of
generating such a state, then its characteristic distribution of pions could
not be regarded as a signal for formation of a quark-gluon plasma. In the
present work we examine whether such states can form during the cooling and
expansion of hot droplets of quark plasma coupled to chiral fields.

Since a prerequisite for DCC formation is that the system should evolve
far from equilibrium, questions of whether a DCC forms and how it evolves
cannot be addressed in the framework of equilibrium thermodynamics but require
either a transport theory or a hydrodynamical approach. Techniques for
applying QCD directly to such situations do not exist at present. Hence most
authors who have investigated this subject have worked within the framework of
the linear sigma model\cite{GML}. Some of them have made approximations that
allow them to construct analytical solutions\cite{BK}. Others have solved the
evolution equations for the chiral fields numerically, using a variety of
approaches and initial
conditions\cite{HW,GGP,GM,AHW,CM,BVH,CKMP,MM,LDC,Ran,KV}. Of these works on
DCC's, the only one that, like ours, includes explicit quark degrees of
freedom is that of Csernai and Mishustin\cite{CM}. Those authors studied
infinite quark matter undergoing a boost-invariant expansion in one direction.
This allowed them to look for the onset of instability with respect to
fluctuations of the chiral fields. In contrast we study here the full
evolution starting from finite-sized droplets of quark matter.

We work within the framework of the linear sigma model\cite{GML}. This is
commonly used as a model for the QCD phase transition because it respects the
SU(2)$\times$SU(2) chiral symmetry of QCD with two light flavours of quark and
it contains a scalar field ($\sigma$) that has the same chiral properties as
the quark condensate. The $\sigma$ field can thus be used to represent the
order parameter for the chiral phase transition. Moreover it has been argued
that the phase transition of the linear sigma model (without quarks) should
lie in the same universality class as that of QCD\cite{Wil} (see
also\cite{RW,Raj}).

Nonetheless universality does not provide a sufficient justification for the
linear sigma model at temperatures that are well below the phase transition or
in situations that are far from thermal equilibrium. The use of the model to
discuss the possibility of forming a DCC thus remains an act of faith, based
on the fact that the model possesses chiral symmetry and contains the most
important degrees of freedom at low energies. If it is extended to include
quarks (and if necessary gluons), the model can be used to describe the
degrees of freedom relevant at high temperatures. At intermediate temperatures
the quarks can also provide a rough model for the effects of massive hadrons
(such as vector mesons and nucleons). An alternative way to include the
effects of these particles would be to add terms involving higher derivatives
of the chiral fields to the Lagrangian, as is done in chiral perturbation
theory\cite{EGPR}. However we do not expect such contact interactions to
provide a good description of the effects of heavy hadrons at the energies of
interest for the question of DCC formation and we believe that inclusion of
explicit quarks provides more appropriate way to extend the model in this
context.

A final comment that we need to make concerns the fact that the quarks in our
model are unconfined and can escape to infinity, whereas in a more realistic
model they should be converted into hadrons. We assume here that the
hadronisation of the fast-moving quarks occurs well outside the original
droplet and so does not affect the subsequent evolution of the chiral fields
inside the droplet.

Our basic approach is very similar to that of Ref.\cite{CM} in that we assume a
rapid quench that leaves the quarks out of thermal equilibrium with the chiral
fields. The subsequent evolution of the system is then described by the
classical Euler-Lagrange equations for the fields coupled to a transport
equation for the quarks. The evolution of the quark density is described by the
relativistic transport equation for fermions in the presence of chiral fields
which has been derived by Shin and Rafelski\cite{SR} and Zhuang and
Heinz\cite{ZH} (see also\cite{ABZH}). Like the authors of Ref.\cite{CM}, we
work in the classical limit where this equation has the form of a relativistic
Vlasov equation for the distribution of the quarks in phase space. The scalar
and pseudoscalar quark densities provide source terms in the Euler-Lagrange
equations for the chiral fields. The classical system of coupled Vlasov and
field equations is solved using a test-particle method~\cite{Won,SG,VBM}.

In using the classical Vlasov equation we are neglecting the nonlocalities
that would be present in a fully quantum mechanical treatment of the quarks.
Our chiral fields do display strong oscillations in both space and time, but
in general they do so in regions where the quark density is small. Hence 
it is unlikely that serious errors are introduced by the classical treatment
of the quarks. We have further neglected collision terms in the transport
equations that could give rise to dissipative effects. Although their
inclusion in a transport theory for quarks poses significant 
problems~\cite{EH}
that go beyond the scope of our current investigation, it is important that
our approach be extended to include collision terms since these could
significantly alter the free streaming of the quarks in our present treatment.

We study here evolution starting from a spherically symmetric droplet of quark
plasma within which the quarks are taken to have a uniform density and a
thermal momentum distribution at some temperature. The $\sigma$ field inside
the droplet is taken to have the value determined self-consistently from the
scalar density of these quarks. We add a small pionic perturbation to this
configuration and then examine whether the pion field grows as the system
evolves. Since our initial conditions do not include a realistic distribution
of thermal pions we cannot estimate what ratio of coherent to incoherent pions
is produced by the decay of such a droplet. Our purpose here is to study
whether such a system can generate coherent pions, and if so, by what
mechanism.

Our results show that the finite size of the system plays a crucial role
in determining its behaviour. In our numerical calculations the quarks rapidly
escape from the original region in which they were placed ({\it
cf}.\cite{AA,GR}). This leaves the chiral fields in an unstable configuration.
We find that the fields always ``roll" away from this configuration in the
$\sigma$ direction, that is towards the true vacuum which surrounds the
droplet. This demonstrates the importance of surface effects. In contrast to
the picture suggested on the basis of bulk systems, there is no indication
that such a droplet can form a region of more-or-less constant fields
corresponding to a misaligned vacuum. Nonetheless we do find that, for large
droplets, there can be a coherent amplification of any initial pion field.
These pion fields display strong oscillations in space and time and so the
behaviour we find is closer to that in Refs.\cite{Ans,RW} than that proposed
in Ref.\cite{Bjo}.

The mechanism that produces this amplification involves the strong
oscillations of the $\sigma$ field that pump energy into oscillations of the
pion field. Similar behaviour has also been seen in Refs.\cite{BVH,MM}. Such a
system is thus more like a ``pion laser" than the traditional picture of a
DCC. A further difference from a DCC is that a chiral phase transition is not
required to produce these enhanced pion fields.

The paper is organized as follows. In Sec.~\ref{tf} we present the 
linear sigma model that we use together with the classical transport and
field equations. The method of solution and initial conditions are described
in Secs.~\ref{method}~and~\ref{ics} respectively. Our results are described in
Sec.~\ref{res}. Their implications are discussed in Sec.~\ref{disc} and ways
in which our approach could be improved are outlined.

\section{Theoretical framework}
\label{tf}

\subsection{Linear sigma model}
\label{model}

A simple model that embodies the main features of QCD associated with chiral
symmetry is the linear sigma model~\cite{GML}. This model respects partial
conservation of the axial current (PCAC) and includes a scalar field that can
model changes in the quark condensate corresponding to restoration of chiral
symmetry. The Lagrangian for the two-flavour version of this model is
\begin{equation}\label{La}
{\cal L} = \bar\psi[i\partial\llap/-g(\sigma+i\mbox{\boldmath$\pi$}\cdot
\mbox{\boldmath$\tau$}\gamma_5)]\psi +\frac{1}{2}(\partial_{\mu}\sigma 
\partial^{\mu}\sigma + \partial_{\mu}\mbox{\boldmath$\pi$}\partial^{\mu}
\mbox{\boldmath$\pi$}) - U(\sigma,\mbox{\boldmath$\pi$}),
\end{equation}
where $\sigma$ and \mbox{\boldmath$\pi$} are the scalar meson and pion fields
mentioned above. Although colour plays no dynamical role in our calculations, 
the quark field $\psi$ that we use describes quarks that come in three colours
as well as two flavours.

The interactions among the meson fields are described by the potential $U$,
which we take to be of the form
\begin{equation}\label{lsm}
U(\sigma,\mbox{\boldmath$\pi$})= \frac{\lambda^2}{4}\bigl(\sigma^2+\pi^2-\nu^2
\bigr)^2-f_{\pi} m_{\pi}^2\sigma,
\end{equation}
where $f_\pi=93$ MeV is the pion decay constant and the parameter $\nu^2$ is 
given by $\nu^2=f_{\pi}^2-m_{\pi}^2/\lambda^2$. Apart from the final term in
$U$ the Lagrangian is symmetric under SU(2)$\times$SU(2) chiral symmetry.
The ``Mexican-hat" form of this potential leads to spontaneous breaking of this
symmetry. In the vacuum the scalar field has a nonzero expectation value
$\sigma=f_\pi$, which corresponds to the quark condensate of the QCD vacuum.
The pions would be massless Goldstone bosons if the final term in $U$ were not
present. In contrast the scalar mesons have a large mass, related to the 
coupling $\lambda^2$ by
\begin{equation}\label{msigma}
m_{\sigma}^2=2\lambda^2 f_{\pi}^2+m_\pi^2~.
\end{equation}

The mass of the $\sigma$ meson is often taken to be around 500 MeV, since the
attractive force between nucleons can be described by exchange of a $\sigma$
meson with such a mass. However one should remember that there are important
contributions from two-pion exchange in this channel and hence the mass of
this ``effective" $\sigma$ meson should not be interpreted as the mass of the
underlying $q\overline q$ state. That state may be better identified with the
$f_0(1370)$ of the particle data tables\cite{PDG}, although Au, Morgan and 
Pennington\cite{AMP} have suggested that this may be part of a much broader
structure in $\pi\pi$ scattering at around 1000 MeV. For most of our work we 
have taken $m_\sigma=1000$ MeV, but we have also considered values in the range
600--1000 MeV.

The quark-meson coupling constant $g$ is more conveniently specified in terms
of the dynamical mass of the quarks in the vacuum with spontaneously broken 
chiral symmetry,
\begin{equation}\label{mq}
M_{\rm q}=gf_{\pi}.
\end{equation} 
The values for $M_{\rm q}$ that we use lie between 300 and 500 MeV,
corresponding to $3.23<g<5.38$.

In our studies we consider configurations in which only one component of the 
pion field is nonvanishing. As a further simplification in our numerical work
we neglect the isospin dependence of the quark-pion coupling. In fact this 
coupling is unimportant for the systems we consider because the quark density 
is close to zero by the time that the pion field becomes important. In practice
therefore we are working with the simplified Lagrangian
\begin{equation}\label{La2}
{\cal L} = \bar\psi[i\partial\llap/-g(\sigma+i\pi\gamma_5)]\psi 
+\frac{1}{2}(\partial_{\mu}\sigma\partial^{\mu}\sigma 
+ \partial_{\mu}\pi\partial^{\mu}\pi) - U(\sigma,\pi),
\end{equation}
in which the meson fields of an O(2) linear sigma model are coupled to two
flavours of quark. The extension of our approach to the O(4) case is 
straightforward.

\subsection{Equations of motion}
\label{eom}

As in the treatment of Csernai and Mishustin\cite{CM}, we assume that the 
expanding droplet of quark matter undergoes a rapid quench which leaves
the quarks out of equilibrium with the chiral fields. We treat the subsequent
evolution of the system classically, ignoring possible effects of quantum
fluctuations.

The classical equations of motion for the meson fields can be derived
straightforwardly from the Lagrangian (\ref{La}). These nonlinear
Klein-Gordon equations take the forms
\begin{equation}\label{sigeq}
\partial^\mu\partial_\mu\sigma= - \lambda^2\bigl(\sigma^2+\pi^2
-\nu^2\bigr)\sigma + f_{\pi} m_{\pi}^2- g \langle\bar{\psi}\psi\rangle,
\end{equation}
\begin{equation}\label{pieq}
\partial^\mu\partial_\mu\pi= - \lambda^2\bigl(\sigma^2+\pi^2-\nu^2
\bigr)\pi - g\langle \bar{\psi}i\gamma_5 \psi\rangle.
\end{equation}

The quarks are described using a relativistic transport theory for fermions in
the presence of scalar and pseudoscalar fields\cite{SR,ZH,ABZH}. We work in
the classical limit (zeroth order in $\hbar$) where the distributions of
quarks and antiquarks in phase space satisfy equations of Vlasov form. The
evolution of the quark distribution in phase space $f(t,{\bf r},{\bf p})$ is
determined by the equation
\begin{equation}\label{eq1}
\displaystyle \left[{\partial\over\partial t} + 
\frac{{\bf p}}{E(t,{\bf r},{\bf p})} \cdot \mbox{\boldmath $\nabla$}_{\bf r} 
-\Bigl(\mbox{\boldmath $\nabla$}_{\bf r} E(t,{\bf r},{\bf p})\Bigr) \cdot 
\mbox{\boldmath $\nabla$}_{\bf p}\right]
f(t,{\bf r},{\bf p}) = 0 ,
\end{equation}
where $E(x,{\bf p})$ is the energy of a relativistic quark at the space-time
point $x=(t,{\bf r})$,
\begin{equation}\label{quarkE}
E(x,{\bf p})=\sqrt{{\bf p}^2 + M^2(x)},
\end{equation}
and $M(x)$ is its mass which is related to the meson fields by
\begin{equation}\label{quarkM}
M(x)=g\sqrt{\sigma^2(x)+\pi^2(x)}.
\end{equation}
The equation for the antiquark distribution, denoted by $\tilde f(t,{\bf
r},{\bf p})$, is identical in form to (\ref{eq1}) since no vector fields are
present in our model.

The Vlasov equation (\ref{eq1}) can be obtained as the classical limit of an
equation for the equal-time Wigner function of the quarks\cite{BGR,ZH1,SR,ZH}.
In fact it describes freely streaming classical quarks and antiquarks, each
obeying the relativistic single-particle equations of motion:
\begin{equation}\label{spe1}
\dot{\bf r}(t)={{\bf p}(t)\over E\Bigl(t,{\bf r}(t),{\bf p}(t)\Bigr)},
\end{equation}
\begin{equation}\label{spe2}
\dot{\bf p}(t)=-\mbox{\boldmath$\nabla$}_{\bf r}
E\Bigl(t,{\bf r}(t),{\bf p}(t)\Bigr),
\end{equation}
where the dots denote time derivatives and the dependence of the particle's
energy on its position and momentum is given by (\ref{quarkE},\ref{quarkM}).
Instead of solving (\ref{eq1}) as a partial differential equation in seven
dimensions, one can replace the smooth distributions $f(t,{\bf r},{\bf p})$
and $\tilde f(t,{\bf r},{\bf p})$ by a set of classical particles obeying
the equations of motion (\ref{spe1},\ref{spe2}). This is the basis for the
test-particle method\cite{Won,SG} which we use to construct approximate
numerical solutions to (\ref{eq1}), as described in the next section.

The couplings of a classical quark to the $\sigma$ and pion fields can
be obtained by differentiating its energy (\ref{quarkE}) with respect to 
each of those fields. The resulting scalar and pseudoscalar quark densities 
are related to the quark and antiquark distributions by
\begin{equation}\label{rhos}
\langle\bar{\psi}\psi(x)\rangle = g\sigma(x) \int {\rm d}^3{\bf p}\, 
\frac{f(x,{\bf p})+\tilde f(x,{\bf p})}{E(x,{\bf p})},
\end{equation}
\begin{equation}\label{rhop}
\langle\bar{\psi}i\gamma_5\psi(x)\rangle = g\pi(x) \int {\rm d}^3{\bf p}\,
\frac{f(x,{\bf p})+\tilde f(x,{\bf p})}{E(x,{\bf p})}.
\end{equation}
These can also be obtained from the classical limit of the Wigner function, as
in Refs.\cite{SR,ZH}. Note that both scalar and pseudoscalar densities vanish
for massless quarks ($\sigma=\pi=0$). Using these expressions in the source
terms for the fields, the field equations (\ref{sigeq},\ref{pieq}) can be
rewritten in the form
\begin{equation}\label{eq3}
\left({\partial^2\over\partial t^2}-\nabla^2\right)\sigma(t,{\bf r})
= - \left[\lambda^2\Bigl(\sigma^2(t,{\bf r})+\pi^2(t,{\bf r})-\nu^2\Bigr) 
+ g^2 \int {\rm d}^3{\bf p}\,
\frac{f(t,{\bf r},{\bf p})+\tilde f(t,{\bf r},{\bf p})}{E(t,{\bf r},{\bf p})} 
\right]\sigma(t,{\bf r})
+ f_{\pi} m_{\pi}^2 ,
\end{equation}
\begin{equation}\label{eq4}
\left({\partial^2\over\partial t^2}-\nabla^2\right)\pi(t,{\bf r})
= - \left[\lambda^2\Bigl(\sigma^2(t,{\bf r})+\pi^2(t,{\bf r})-\nu^2\Bigr) 
+ g^2 \int {\rm d}^3{\bf p}\,
\frac{f(t,{\bf r},{\bf p})+\tilde f(t,{\bf r},{\bf p})}{E(t,{\bf r},{\bf p})} 
\right]\pi(t,{\bf r}).
\end{equation}
The evolution of our system is obtained by solving self-consistently the
set of differential equations (\ref{eq1},\ref{eq3},\ref{eq4}) subject to
an appropriate set of initial conditions (to be described in Sec.~\ref{ics}).

\section{Method of solution}
\label{method}

Solutions of the set of coupled equations (\ref{eq1},\ref{eq3},\ref{eq4}) 
can be obtained only numerically. The methods we adopt are very similar to 
those employed in Ref.\cite{VBM} to solve the similar set of equations that
arise in a soliton bag model. 

The Vlasov equation (\ref{eq1}) is conveniently solved using the test-particle
method which has been widely used in the context of intermediate-energy
heavy-ion collisions\cite{Won,SG}, $NN$ scattering \cite{VBM} and evolution
of a quark plasma\cite{AA}. The method consists of replacing the quark and 
antiquark distributions by a swarm of test particles. The smooth distributions
$f(t,{\bf r},{\bf p})$ and $\tilde f(t,{\bf r},{\bf p})$ are thus approximated
by
\begin{equation}\label{f} 
{\displaystyle f (t, {\bf r},{\bf p}) = w \sum_{n=1}^{N} 
\delta^3 ({\bf r}-{\bf r}_n(t)) ~\delta^3({\bf p}-{\bf p}_n(t)) } 
\end{equation}
\begin{equation}\label{ft} 
{\displaystyle \tilde f (t, {\bf r},{\bf p}) = w \sum_{n=1}^{\tilde N} 
\delta^3 ({\bf r}-\tilde {\bf r}_n(t)) ~\delta^3({\bf p}-\tilde {\bf p}_n(t))}
\end{equation}
Each test particle follows a classical trajectory ${\bf r}_n(t)$, ${\bf
p}_n(t)$ determined by the relativistic single-particle equations of motion
\begin{equation}
\dot{\bf r}_n(t)={{\bf p}_n(t)\over E\Bigl(t,{\bf r}_n(t),{\bf p}_n(t)\Bigr)},
\end{equation}
\begin{equation}
\dot{\bf p}_n(t)=-\mbox{\boldmath$\nabla$}_{\bf r}
E\Bigl(t,{\bf r}_n(t),{\bf p}_n(t)\Bigr).
\end{equation}
Provided enough test particles are used, their distributions provide very good
approximations to the smooth quark and antiquark distributions described by
(\ref{eq1}). In general the number of test particles is very much larger than
the actual numbers of quarks and antiquarks we wish to describe. To account
for this, the distributions must be multiplied by a normalization constant $w$.
If the actual numbers of quarks and antiquarks are denoted by $A$ and $\tilde 
A$ respectively then the numbers of test particles are related to these by
\begin{equation}
A=\int{\rm d}^3{\bf r}\,{\rm d}^3{\bf p}\,f(t,{\bf r},{\bf p}) = wN,
\end{equation}
\begin{equation}
\tilde A=\int{\rm d}^3{\bf r}\,{\rm d}^3{\bf p}\,\tilde f(t,{\bf r},{\bf p}) 
= w\tilde N.
\end{equation}

The nonlinear equations for the meson fields (\ref{eq3},\ref{eq4}) are 
second-order in both space and time derivatives. To solve these we work on a
discrete mesh of points in space. In the present work we confine our attention 
to spherically symmetric initial conditions. We further require that the 
symmetry is maintained during the evolution, by angle-averaging the source
terms in the field equations. In this case, the field equations can be written
\begin{equation}\label{sig}
\displaystyle
\ddot\sigma=\frac{1}{r}\frac{\partial^2}{\partial r^2}(r\sigma)
-\left[\lambda^2\Bigl(\sigma^2(t,r)+\pi^2(t,r)-\nu^2\Bigr) 
+ g^2 {\cal S}_{\rm q}(t,r)\right]\sigma(t,r) + f_{\pi} m_{\pi}^2,
\end{equation}
\begin{equation}\label{pi}
\displaystyle
\ddot\pi=\frac{1}{r}\frac{\partial^2}{\partial r^2}(r\pi)
-\left[\lambda^2\Bigl(\sigma^2(t,r)+\pi^2(t,r)-\nu^2\Bigr) 
+ g^2 {\cal S}_{\rm q}(t,r)\right]\pi(t,r),
\end{equation}
where the angle-averaged source term is
\begin{equation}\label{Sq}
{\cal S}_{\rm q}(t,r)=\frac{w}{4\pi}\left(
\sum_{n=1}^N \frac{1}{r_n^2E_n}\delta(r-r_n(t))+
\sum_{n=1}^{\tilde N} \frac{1}{\tilde r_n^2\tilde E_n}\delta(r-\tilde r_n(t))
\right).
\end{equation}
The restriction to spherically symmetric configurations considerably lessens
the computational effort required to solve these equations. In addition
the angle average in (\ref{Sq}) reduces the number of test particles required
to accurately map out the solution of the Vlasov equation. As a check on the
consistency of this restriction we have run some small-scale three-dimensional
simulations and have found no evidence for instability against non-spherical
perturbations.

The field equations are solved on a radial mesh with spacing  $\Delta
r=0.05$~fm. In most of our runs we have used a mesh of 280 points,
corresponding to a region of radius $R=14$ fm. At the outer edge of the
lattice we impose the boundary condition that the pion field vanishes while
the $\sigma$ field is set equal to $f_{\pi}$. This means that we do not lose
energy from these fields when the radiated meson waves reach the edge, which is
useful since we wish to calculated the energy radiated in particular modes of
the pion field. However the fact that meson waves are reflected by the
boundary means that our results become unphysical once those reflected waves
return to the central region. For an initial droplet of size $r_0=4$ fm and a
lattice of size $R=14$ fm this occurs after about 18--20 fm/$c$. To describe
evolution over longer times we need to use lattices with larger radii or to
work with absorbing boundary conditions.

The discrete nature of the source term due to the test particles requires
that the $\delta$-functions in (\ref{Sq}) be smeared out in some way. In most 
of our calculations we have used a parabolic form,
\begin{equation}
\label{par}
\delta(r-r_n)\rightarrow {3\over 4d^3}\Bigl(d^2-(r-r_n)^2\Bigr),\qquad 
|r-r_n|\leq d.
\end{equation}
In order that all test particles contribute to the source, the thickness $d$
should be greater than half the mesh size: $d\ge \Delta r/2$. In practice $d$
should be somewhat larger than this to ensure a reasonably smooth source. The
results shown in this paper correspond to $d=2 \Delta r$. We have checked that
our results are not sensitive to the precise value of $d$ in this region, nor
do they change if Eq.~(\ref{par}) is replaced by a Gaussian smearing function.

For the initial conditions that we study we use an initial density of test
particles of about 400--500 fm$^{-3}$. For a region of radius 4 fm this
requires a total of about 125000 test particles. We have checked that our
results are stable with respect to a further increase in the number of test
particles.

The coupled equations are integrated using a leapfrog method with a time step
of $\Delta t=0.01$ fm/$c$. We have also  compared our method with the more
complicated staggered leapfrog method\cite{HK} described in the Appendix of
Ref.~\cite{VBM}, and checked that the two agree. Since the coupled equations
corresponds to a conservative system, conservation of the total energy
provides a good test of the accuracy of our numerical algorithm. We find that
the energy changes by less than 1\% over a time of 40 fm/$c$ (4000 time steps).

\section{Initial conditions}
\label{ics}

Our initial configuration consists of a spherical droplet of hot quark matter.
We take the quark density and temperature to be uniform within the droplet of
radius $r_0$. The initial distribution of the quarks in phase space is thus
\begin{equation}\label{finit}
f(0,{\bf r}, {\bf p})={2N_c N_f\over (2\pi)^3}\Theta(r_0-r)\,
\eta\Bigl(E_0({\bf p}),T,\mu\Bigr)
\end{equation}
where $N_c=3$ and $N_f=2$ are the numbers of colours and flavours, 
respectively. $\Theta$ denotes the step function and $\eta(E,T,\mu)$ 
is the Fermi distribution
\begin{equation}\label{fermi}
\eta(E,T,\mu)={1 \over 1+e^{(E- \mu)/T}}.
\end{equation}
For the antiquark distribution, $\eta$ should be replaced by
\begin{equation}\label{fermit}
\tilde \eta (E,T,\mu)= \eta (E,T,-\mu).
\end{equation}
The quark energy appearing in these distributions has the form $E_0({\bf p})=
\sqrt{{\bf p}^{2} + M_0^2}$ where the quark mass $M_0$ is obtained 
by solving self-consistently the equations for the constant $\sigma$ field in
the presence of the (anti-)quark distributions (\ref{fermi},\ref{fermit}). For
a uniform quark density the equation (\ref{sigeq}) for the $\sigma$ field
takes the form
\begin{equation} \label{gap}
\displaystyle
\left[\lambda^2\Bigl(\sigma_0^2-\nu^2\Bigr)+g^2\frac{2N_cN_f}{(2\pi)^3} 
\int\!{\rm d}^3{\bf p}\,\frac{\eta(E_0,T,\mu)+\tilde \eta(E_0,T,\mu)}{E_0} 
\right]\sigma_0-f_{\pi} m_{\pi}^2=0,
\end{equation}
where $E_0({\bf p})$ implicitly depends on $\sigma_0$ since $M_0=g\sigma_0$.
In the chiral limit ($m_\pi^2=0$) this equation always has a trivial solution 
$\sigma_0=0$, corresponding to a vacuum in which chiral symmetry is restored.
For high enough temperatures or densities this is the only solution to
(\ref{gap}). At low temperature and density there is also a
nontrivial solution with lower energy which describes a vacuum in which chiral
symmetry is spontaneously broken.

The temperature at which the solution with $\sigma_0\neq 0$ becomes the ground
state (for fixed $\mu$) acts like a critical temperature. For the parameter
set $m_\sigma=1000$ MeV, $M_{\rm q}=300$ MeV this temperature is $T_0\simeq235$
MeV, while for $m_\sigma=600$ MeV we find $T_0\simeq150$ MeV. Of course, one
should remember that thermal fluctuations of the meson fields have been
ignored in our treatment and so $T_0$ is somewhat higher than the true
critical temperature for the model. Nonetheless we can make use of the
existence of this temperature to explore initial states in which chiral
symmetry is restored as well as ones in which it is spontaneously broken. If
chiral symmetry is explicitly broken then there is no phase transition.
Instead one has a smooth crossover from states with $\sigma_0$ close to $f_\pi$
to ones with small, but still nonzero, values of $\sigma_0$. Nonetheless, for
a symmetry breaking strength that gives a realistic pion mass, this crossover
behaviour occurs rapidly and one can still usefully define a critical
temperature above which the approximate chiral symmetry is restored and the
quarks are nearly massless.

For the initial $\sigma$ field, we pick a form that interpolates smoothly
between $\sigma_0$ inside the droplet and $f_\pi$ outside:
\begin{equation} \label{sig0}
\sigma(t=0,r)= \sigma_0 +(f_\pi-\sigma_0)\Theta_\alpha(r-r_0),
\end{equation}
where $\Theta_\alpha$ is a suitably smoothed step function. In this work we
have chosen to take
\begin{equation}\label{step}
\Theta_{\alpha}(r-r_0) =\left\{ 
\begin{array}{ll} 
&\displaystyle (e^{ar^2}-1)/2(e^{ar_0^2}-1) ~~~~~~~r< r_0 \\
&\displaystyle (1+\tanh\alpha(r-r_0)~)/2 ~~~~~r\ge r_0 ,
\end{array} \right.
\end{equation}
where $a$ is determined by requiring the radial derivative of
$\Theta_{\alpha}$ to be continuous at $r=r_0$. This is a convenient 
parametrisation whose shape is close to the self-consistent solution of the 
field equation for $\sigma$ in the presence of a static spherical quark 
distribution of the form (\ref{finit}). The constant $\alpha$ which describes
the inverse of the surface thickness should be proportional to $m_{\sigma}$.

We take the initial pion field to be of the form
\begin{equation}\label{pi0}
\displaystyle
\pi(t=0,r)= \pi_0\, \Bigl(1-\Theta_{\alpha}(r-r_0)\Bigr),
\end{equation}
where $\pi_0$ is a small, arbitrary amplitude corresponding to a fluctuation
away from the self-consistent solution of (\ref{gap}). A nonzero initial value
for either the pion field or $\dot\pi(0,r)$ is needed since otherwise our
system will never develop a pion field, as can be seen from the fact that all
the terms in the field equation (\ref{pi}) contain a factor of $\pi(t,r)$.
Ultimately we would hope to take the initial fluctuations in both $\sigma$ and
pion fields from a thermal distribution, using an extension of the method
described in\cite{Ran2}. However for our current investigations we use the
simple ansatz (\ref{sig0},\ref{pi0}) in order to examine whether a system of
this type can lead to a coherent pion field.

The equations of motion of the meson fields (\ref{sig},\ref{pi}) are of 
second order in time derivatives and so we need to specify the initial field
velocities as well as the fields. In most of our calculations we have used
static initial configurations, $\dot\sigma(0,r)=\dot\pi(0,r)=0$. However we have
also considered more general cases, and in particular ones corresponding to 
droplets that are expanding with velocity $v$:
\begin{equation}\label{sigv0}
\dot \sigma(0,r)=-v {\partial\sigma\over\partial r}(0,r),
\end{equation}
\begin{equation}\label{piv0}
\dot \pi(0,r)=-v{\partial\pi\over\partial r}(0,r).
\end{equation}

\section{Results}
\label{res}

In this section, we present the results of our simulations corresponding
to various initial conditions of the system. Most of our results are for
our ``standard" parameter set, $m_{\sigma}=1000$ MeV,  $m_{\pi}= 139$ MeV and 
$M_{\rm q}=300$ MeV. For these $\sigma$ and quark masses, in the chiral limit,
chiral symmetry restoration occurs at the temperature $T_0\simeq 235$ MeV, as
discussed above. 

For these parameters we have examined the evolution of droplets of various
sizes, with an initial temperature of 250 MeV (above $T_0$) and zero chemical
potential (zero baryon density). The initial energy density in this case is
about 3.5 GeV/fm$^{-3}$, comparable to the energy densities that are expected
to be reached in ultrarelativistic heavy-ion collisions. This energy density
is dominantly due to the quarks, the chiral fields providing only about 0.14
GeV/fm$^{-3}$. Inside the droplets the initial value of the sigma
field, determined from the gap equation (\ref{gap}) with explicit symmetry
breaking, is $\sigma_0=10$ MeV. In order to study whether initial pionic
fields can be amplified during the evolution we add a pionic perturbation to
this configuration of magnitude $\pi_0=0.05 f_{\pi}$.

We examine first the evolution of a small initial droplet of radius $r_0=1.1$
fm, plotting in Fig.~1 the $\sigma$ and pion fields at several radii as a
function of time. This shows the fields rolling away from the initial
configuration towards the physical vacuum surrounding the initial droplet.
Both fields then oscillate around their vacuum values, their amplitudes
decaying as energy is carried away (mainly by long-wavelength sigma waves).
Fig.~1 shows no indication of any enhancement of the pion field. We find
similar results for small droplets with $r_0<2$ fm.

In contrast, for larger initial droplets we find significant amplification of
the pion field, with strong oscillations in space and time. This is shown in
Fig.~2 by the time evolution of the $\sigma$ and pion fields in the case of a
droplet of radius $r_0=4.0$ fm, with otherwise the same initial conditions as
in Fig.~1. (The fluctuations that can be seen in the fields at $r=0.1$ fm for
$t<4$ fm/$c$ are numerical noise, due to the fact that the radial density of
test quarks is much smaller at small radii.) Another view of the same system
is provided in Fig.~3, where we show the $\sigma$ and pion fields as a
function of $r$ at successive times. A larger initial volume means that the
energy stored initially in the meson fields is larger, and this initial energy
is crucial to determining whether or not the pion field is amplified.

Also from Fig.~2 we see that the strong oscillations of the fields start only
after 4--6 fm$/c$. This is the time needed by the quarks to escape the initial
region. Indeed we find that for our standard choice of parameters almost all
the quarks rapidly stream out of the initial region. This leaves the chiral
fields in an unstable configuration which collapses towards the true vacuum.
This behaviour can be seen in Fig.~4 where we plot the net density of quarks
and antiquarks: $\int\!{\rm d}^3{\bf p}\,(f(t,{\bf r},{\bf p})+\tilde f(t,{\bf
r},{\bf p}))$ times $r^2$ as a function of $r$ at successive times. 

In order to investigate the spectrum of the pion modes that are excited we 
Fourier analyse the pion field at successive times. From the spatial Fourier
transform of the pion field, $\tilde \pi(t,{\bf k})$, and its time derivative,
$\tilde {\dot \pi}(t,{\bf k})$, we calculate the corresponding intensity in
momentum space:
\begin{equation} \label{Epi}
{\cal E}_{\pi}(t,{\bf k})={\textstyle{1\over 2}}\Bigl(\vert\dot{\tilde\pi}
(t,{\bf k})\vert^2 + \omega_k^2 \vert\tilde \pi(t,{\bf k})\vert^2\Bigr),
\end{equation}
where $\omega_k=\sqrt{k^2+m_\pi^2}$. If the pion fields are sufficiently weak
that they are well described by the linearised version of the equation of
motion (\ref{eq4}), then (\ref{Epi}) is simply the energy density of the pion
field in momentum space. At large times it thus gives the energy radiated in
the form of pions as a function of momentum. The total energy of the pion
field in such a regime is
\begin{equation} \label{Epitot}
E_\pi(t)=\int\!\frac{{\rm d}^3 {\bf k}}{(2\pi)^3}\, {\cal E}_{\pi}(t,{\bf k}),
\end{equation}
and the corresponding number of pions is given by
\begin{equation}\label{Npi}
N_{\pi}(t)= \int\!\frac{{\rm d}^3 {\bf k}}{(2\pi)^3}\,{1\over \omega_k}
 {\cal E}_{\pi}(t,{\bf k}).
\end{equation}
However, one should note that while the system is still evolving nonlinearly,
(\ref{Epi}) has no such interpretation. 

Fig.~5 shows the behaviour of the intensity (\ref{Epi}) at successive times as
a function of momentum. (These momenta are the discrete values
$k_n=n\pi/R$, $n\ge 1$, consistent with our boundary condition at $r=R$.) One
can see from this plot that during the period when the $\sigma$ field is
undergoing violent oscillations (4--8 fm/$c$) the pion modes with momenta
$k\sim 0.5$--2 fm$^{-1}$ are significantly amplified. There is also a
smaller amplification for modes with larger momenta, up to $\sim 4$ fm$^{-1}$.
This enhancement of pion modes with momenta less than $\sim 2$ fm$^{-1}$ is
very similar to the behaviour found by Randrup\cite{Ran}, in spite of the use
of a different approach.

In Fig.~6 we plot ${\cal E}_{\pi}(t,{\bf k})$ for the most strongly amplified
modes as a function of time. This shows that very little further energy is
radiated in these modes after 8 fm/$c$. Similarly the total number of pions,
calculated using (\ref{Npi}) and plotted in Fig.~7, shows very little change
after this time. In this case the number of pions produced is 20 times larger
than the number corresponding to the initial pionic perturbation.

Another related way to study the pionic modes is provided by the Fourier
transform of the correlation function, whose time dependence was plotted by
the authors of Refs.~\cite{RW,Raj}. This quantity is equal to $\vert\tilde
\pi(t,{\bf k})\vert^2$ and so can be thought of as the second term of
(\ref{Epi}) divided by $\omega_k^2$. Once the pion field is in the regime
where it satisfies a linear equation and oscillates harmonically, $\vert\tilde
\pi(t,{\bf k})\vert^2$ oscillates with frequency $2\omega_k$. These
oscillations are smoothed out in~(\ref{Epi}) by the presence of the
time-derivative term. Although we do not present results for the correlation
function here, we have looked at it to check that the enhanced modes shown in
Fig.~6 do indeed show the expected oscillations for radiated pions which are in
the harmonic regime. The small remaining oscillations of the pion field in the
central region that can be seen at times after 8 fm/$c$ in Fig.~2 do not
contribute significantly to the pion radiation. This can also be seen from the
behaviour of the total number of pions in Fig.~7.

To explore the robustness of this mechanism for generating enhanced pion
fields we have considered other initial conditions. These include examples
with nonzero baryon density ($\mu \ne 0$) or expanding initial droplets ($v\ne
0$ in Eqs.~(\ref{sigv0},\ref{piv0})) and initial pionic perturbations with
different magnitudes or spatial extents. In all these cases we obtained
results that are qualitatively similar to those shown here. They are also
qualitatively the same if we work in the chiral limit with massless pions. Our
results are however sensitive to other parameters of the model. In particular
smaller $\sigma$ masses correspond to smaller initial energy densities for the
chiral fields. For example, with $m_\sigma=600$ MeV a droplet of radius 4 fm
leads to much less amplification than the case shown in Figs.~2--7. The effect
of taking a larger quark-meson coupling is discussed below.

All of the cases discussed so far correspond to evolution starting from the
phase with restored (approximate) chiral symmetry. One can also ask about
evolution from an initial state where chiral symmetry is already spontaneously
broken. For this purpose we have considered initial conditions where the
temperature is $T=200$ MeV ($<T_0$) but with other parameters as above. In
this case, the initial hot quarks are ``dressed'' and have a dynamical mass of
$g\sigma_0=206$ MeV where $\sigma_0$ is the nontrivial solution of
(\ref{gap}). We take the radius of the initial droplet and the initial
pionic perturbation to be the same as in the previous example. The time
evolution of the fields at several radii are shown in Fig.~8 for this case.
The qualitative behaviour is similar to that of the example with $T>T_0$ 
shown in Fig.~2. Again we see that $\sigma$ field oscillates strongly
although for a rather shorter period: in this case 4--6 fm/$c$. During these
oscillations there is a significant amplification of the pion field.
This shows that there is nothing about this mechanism for generating a coherent
pion field that requires a phase transition. What is essential is that the
initial chiral fields have sufficient energy.

Csernai and Mishustin\cite{CM} have used the same model to study the simpler
case of infinite quark matter undergoing a boost-invariant expansion in one
direction. They found a similar coherent amplification of the pion field.
In addition they pointed out that a more complete treatment of the model would 
be expected to lead to formation of clusters of quarks and antiquarks
surrounded by domains of the coherent chiral fields. In the cases discussed
above we did not find any such clusters ({\it cf}. Fig.~4). However if we
increase the strength of the quark-meson coupling we do obtain such behaviour.
This can be illustrated by the parameter set $M_{\rm q}=500$ MeV ($g=5.38$) and
$m_{\sigma}=1$~GeV, in the chiral limit. In this case symmetry restoration
occurs at $T_0\simeq 170$ MeV. We consider a droplet of radius $r_0=2.75$ fm,
with initial temperature $T=250$ MeV and chemical potential $\mu=100$ MeV. In
Fig.~9 we show the radial dependence of the density of quarks and antiquarks
times $r^2$ at successive times. In contrast to the case shown in Fig.~4, the
quarks with smaller momenta are reflected from the surface region at around 3
fm and remain trapped within the droplet.

The behaviour of the chiral fields shown in Fig.~10 is quite different from
the previous examples. The $\sigma$ field in the central region continues to
oscillate wildly around values that are much less than $f_\pi$. The quark 
cluster is obviously created in a highly excited configuration and as it
settles down a significant fraction of its excitation energy is radiated in the
form of pions, as can be seen from the fact that very large amplitude
oscillations of the pion field continue for times up to 20 fm/$c$ (and well
beyond). In this particular example the quarks are left in a shell-like
distribution with a radius $\sim 2.5$ fm, which is basically a ``SLAC bag"
soliton\cite{SLAC}. Other initial conditions that we have studied lead to more
uniform distributions of quarks trapped inside shallow ``bags" in the $\sigma$
field. In all cases the resulting excited quark clusters form efficient
radiators of coherent pions.

\section{Summary and Discussion}
\label{disc}

We have investigated the expansion of hot droplets of quark plasma coupled to
the sigma and the pion fields of the linear sigma model, addressing in
particular the question of whether such droplets can lead to the production of
a coherent classical pion field or a disoriented chiral condensate (DCC). It
has been suggested that the formation of a DCC could be used a signal for the
chiral phase transition in high-energy nucleus-nucleus or hadron
collisions\cite{Bjo,RW,Raj}. We have therefore examined whether the phase
transition is the only mechanism that could produce a coherent pion~field.

We treat the evolution of the system classically and solve numerically the
Vlasov equation for the quarks coupled to the nonlinear Klein-Gordon equations
for the chiral fields. These equations have previously been studied by Csernai 
and Mishustin\cite{CM}, but only for the case of infinite quark matter 
undergoing a boost-invariant expansion. Our results show that finite-size
effects play an important role in the evolution of such systems.

Our starting point is a spherical droplet containing a uniform density of 
quarks and antiquarks in thermal equilibrium at some temperature $T$. Inside
this region the sigma field is taken to be constant. In using this type of
initial configuration we are assuming a rapid quench that leaves the quarks
out of thermal equilibrium with the meson fields. We add to this a small pionic
perturbation, to act as a ``seed'' for possible formation of a coherent pion
field. The fluctuations of the initial meson fields away from their
self-consistent values ought to be taken from a thermal distribution. We plan
to do this in future work using an extension of the method proposed by
Randrup\cite{Ran2}. In the present paper we have used a uniform initial
perturbation in order to study whether such droplets can lead to
amplification of a coherent pion field.

We find that the quarks rapidly escape from the original region in which they
were placed, as in Ref.\cite{AA,GR}. This leaves the chiral fields in an
unstable configuration. Starting from the surface of the droplet, the chiral
fields relax towards the physical vacuum which surrounds the droplet
($\sigma=f_{\pi}, \pi=0$). The behaviour is rather different from the picture
which has been suggested on the basis of a uniform system, namely, that the
symmetry of the potential could allow different regions to relax towards
vacuum configurations with different chiral orientations. In the finite
systems that we have studied, the surface effects ensure that the fields
always ``roll'' away from their unstable initial configuration towards the
physical vacuum.

The departure of the quarks leaves the $\sigma$ field in a configuration with
significant potential energy. As a result it undergoes rather violent 
oscillations, during which this energy is radiated in the form of meson waves,
before settling down to its vacuum value. If an initial pion field is present
then some of this energy can be converted into pions. This leads to a coherent
production of pions with a particular isospin and so would have similar
observable effects to a DCC. However unlike the picture of a DCC consisting
regions of differently oriented vacuum, the pion fields oscillate strongly in
space and time. Hence the behaviour we find is closer to that in
Refs.~\cite{Ans,RW,BVH,MM} than that proposed in Ref.~\cite{Bjo}. For large
initial droplets that start in the chirally restored phase, we find
significant amplification of the number of pions, typically by factors of the
order of 20 or more. The mechanism is robust against changes in the initial
conditions as described in Sec.~\ref{res}.

In some cases we find that the pion oscillations contain a strong component
with half the frequency of the $\sigma$ oscillations. Similar oscillations of
pion fields driven by the $\sigma$ field have also been seen in the
homogeneous systems of $\sigma$ and pion fields (without quarks) studied by
Boyanovsky {\it et al.}\cite{BVH} and Mrowczynski and M\"uller\cite{MM}.
Analogous behaviour is also seen in models of inflationary universes where an
additional scalar field is coupled to the field that drives the
inflation\cite{STB,BVHS}. Mathematically one can describe this as a parametric
resonance driven by the oscillations of the $\sigma$-dependent mass term in
the pion field equation. More physically one can view it as a ``pion laser"
where energy is pumped into the pion field by the process
$\sigma\rightarrow\pi\pi$. The larger the initial energy in the $\sigma$
field, the more effectively this mechanism operates. Hence we find the greatest
amplification for the largest droplets, in cases with the largest $\sigma$
mass. Smaller droplets, with radii less than 2 fm, do not lead to significant
pion production. Also the enhancement factors are much smaller for cases with
$m_\sigma=600$ MeV compared to those where we took $m_\sigma=1000$ MeV.

If the quark-meson coupling is taken to be strong enough, then some of the
quarks remain trapped inside the droplet and eventually form a cold multiquark
cluster, as suggested by Csernai and Mishustin\cite{CM}. These clusters are
produced in highly excited configurations that can act as extremely efficient
radiators of coherent pions. However one should remember that the cold quark
matter of the linear sigma model is not expected to be a very good
approximation to hadronic matter. Hence one should not take this version of
the mechanism for coherent pion production too seriously before more realistic
models have been explored.

The nature of the mechanism for the enhancement of the pion field found in our
work makes no obvious reference to the chiral phase transition. As explained
in Sec.~\ref{ics} the model possesses a temperature $T_0$ below which chiral
symmetry is spontaneously broken. This enables us to study evolution from
states in which the symmetry is partially restored, that is, where the
temperature, although less than $T_0$, is still high enough that the $\sigma$
field is significantly reduced from its vacuum value. The matter in these
states consists of dressed quarks with finite dynamical masses, and can be
though of as a crude model for hot hadronic matter in which hadrons with
higher masses than pions are present. We find that large enough droplets with
initial temperatures below $T_0$, behave in a similar way to ones with
initially restored chiral symmetry. In particular they also lead to
amplification of coherent pions. This is a further indication that the
mechanism involved is that of a ``pion laser" which is driven by the
oscillating $\sigma$ field and which does not rely on the phase transition for
its operation.

Like the DCC's and other coherent pion fields discussed in the literature, the
pions produced by this mechanism will have the same characteristic $1/\sqrt f$
distribution for the fraction $f$ of neutral pions. However to determine 
whether this signal will be experimentally observable over the background of
incoherently produced pions we need to know the actual number of coherent
pions. This will require the use of more realistic inital conditions. In
particular, as already mentioned, we need to take the initial pionic
fluctuations from a thermal distribution. For this purpose we will need to
extend the method described in\cite{Ran2} to our model which has explicit quark
degrees of freedom. In addition we are studying other geometries since,
especially for ultrarelativistic collisions, a cylindrical system undergoing a
boost-invariant expansion is expected to be more relevant than a spherical one.
Finally we should also point out the need to extend our approach to include
both quantum effects and collision terms in the transport equations.


\acknowledgements{We are indebted to T. D. Cohen and J.~A. McGovern for many
helpful discussions, and to them and K. Rajagopal for critically reading the
manuscript. A.A.\ is grateful to the ECT*, Trento for its hospitality during
the workshop on DCC's and to many of the participants at that workshop for
useful discussions. Support from the EPSRC and PPARC is acknowledged.}

\vskip 1cm
\begin{center} { {\Large {\bf Figure captions} } }\end{center}

\noindent
{\bf Fig.~1}~: The $\sigma$ (a) and pion (b) fields as functions of time for
various radii. The parameters of the model are: $m_{\sigma}=1$~GeV, $m_{\pi}=
139$~MeV, and $M_{\rm q}=300$~MeV. The initial conditions are: $T=250$~MeV,
$\mu=0$, and $r_0=1.1$~fm.

\vskip 0.2cm \noindent
{\bf Fig.~2}~: As in Fig.~1 except for the initial radius: $r_0=4$~fm.

\vskip 0.2cm \noindent
{\bf Fig.~3}~: The $\sigma$ (a,b) and pion (c,d) fields as functions of radius
at successive times. The parameters of the model as well as the initial
conditions are the same as in Fig.~2.

\vskip 0.2cm \noindent
{\bf Fig.~4}~: The net density of quarks and antiquarks weighted with $r^2$ as
a function of radius. The parameters of the model as well as the initial
conditions are the same as in Fig.~2.

\vskip 0.2cm \noindent
{\bf Fig.~5}~: The energy density of the pion field in momentum space ${\cal
E}_{\pi}$ as a function of momentum at successive times. The parameters of the
model as well as the initial conditions are the same as in Fig.~2.

\vskip 0.2cm \noindent
{\bf Fig.~6}~: The time evolution of ${\cal E}_{\pi}$ for the relevant modes.
The parameters of the model as well as the initial conditions are the same as
in Fig.~2.

\vskip 0.2cm \noindent
{\bf Fig.~7}~: The number of pions as defined in Eq.~(\ref{Npi}) as a function
of time. The parameters of the model as well as the initial conditions are the
same as in Fig.~2.

\vskip 0.2cm \noindent
{\bf Fig.~8}~: The $\sigma$ (a) and pion (b) fields as functions of time for
various radii. The parameters of the model are the same as in Fig.~1 while the
initial conditions are: $T=200$~MeV, $\mu=0$, $r_0=4$~fm.

\vskip 0.2cm \noindent
{\bf Fig.~9}~: The net density of quarks and antiquarks weighted with $r^2$ as
a function of radius, in the case where $m_{\sigma}=1$~GeV,  $m_{\pi}= 0$, and
$M_{\rm q}=500$~MeV. The initial conditions are: $T=250$~MeV, $\mu=100$, and
$r_0=2.75$~fm.

\vskip 0.2cm \noindent
{\bf Fig.~10}~: The $\sigma$ (a) and pion (b) fields as functions of time for
various radii. The parameters of the model as well as the the initial
conditions are the same as in Fig.~9.

\newpage
{\Large {\bf Abada/Birse, Fig.~1, part 1 of 2.} }
\vskip 1 cm
\begin{figure}[t]
\centerline{\psfig{figure=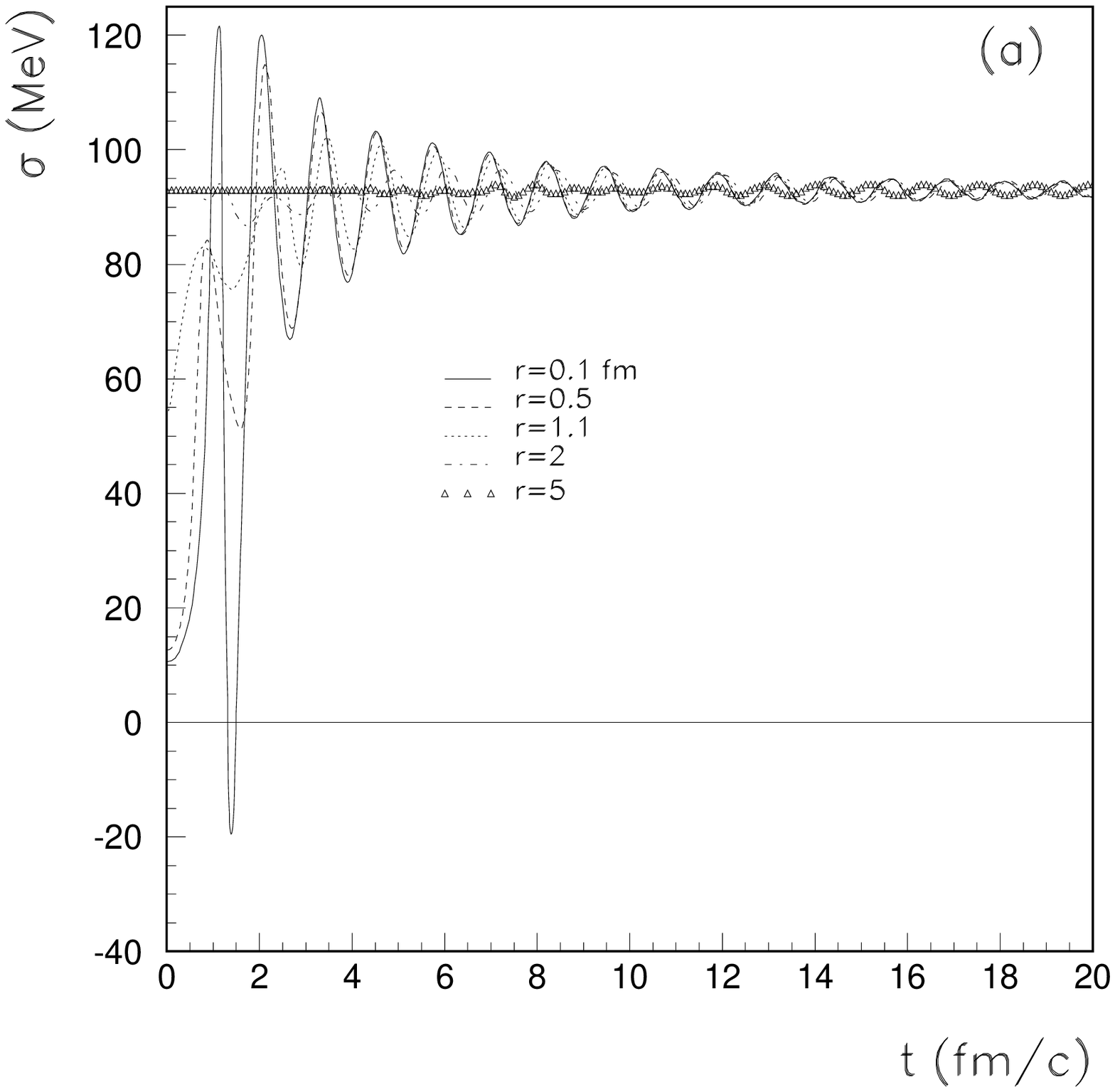}}
\end{figure}

\newpage
{\Large {\bf Abada/Birse, Fig.~1, part 2 of 2.} }
\vskip 1 cm
\begin{figure}[t]
\centerline{\psfig{figure=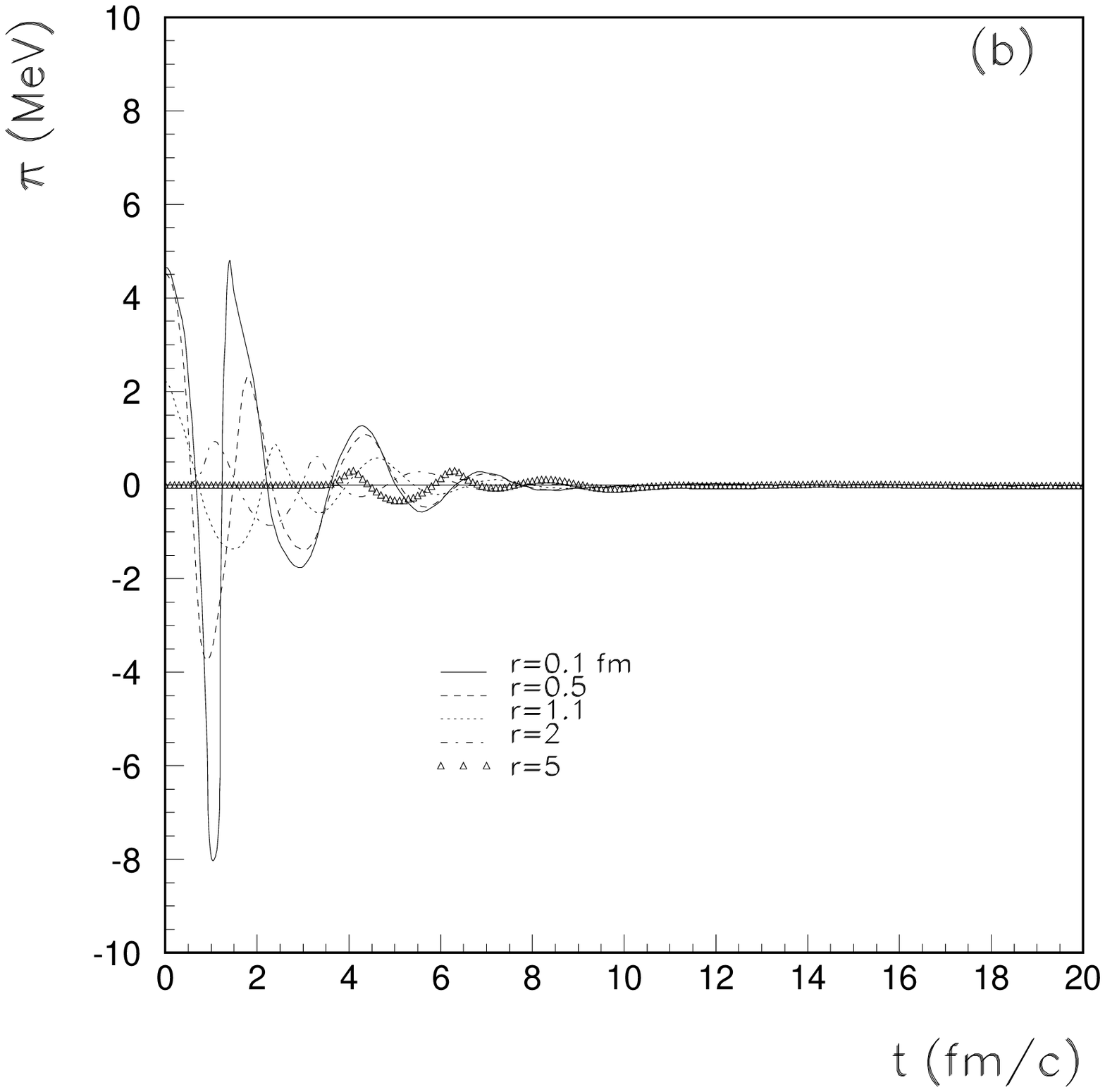}}
\end{figure}

\newpage
{\Large {\bf Abada/Birse, Fig.~2, part 1 of 2.} }
\vskip 1 cm
\begin{figure}[t]
\centerline{\psfig{figure=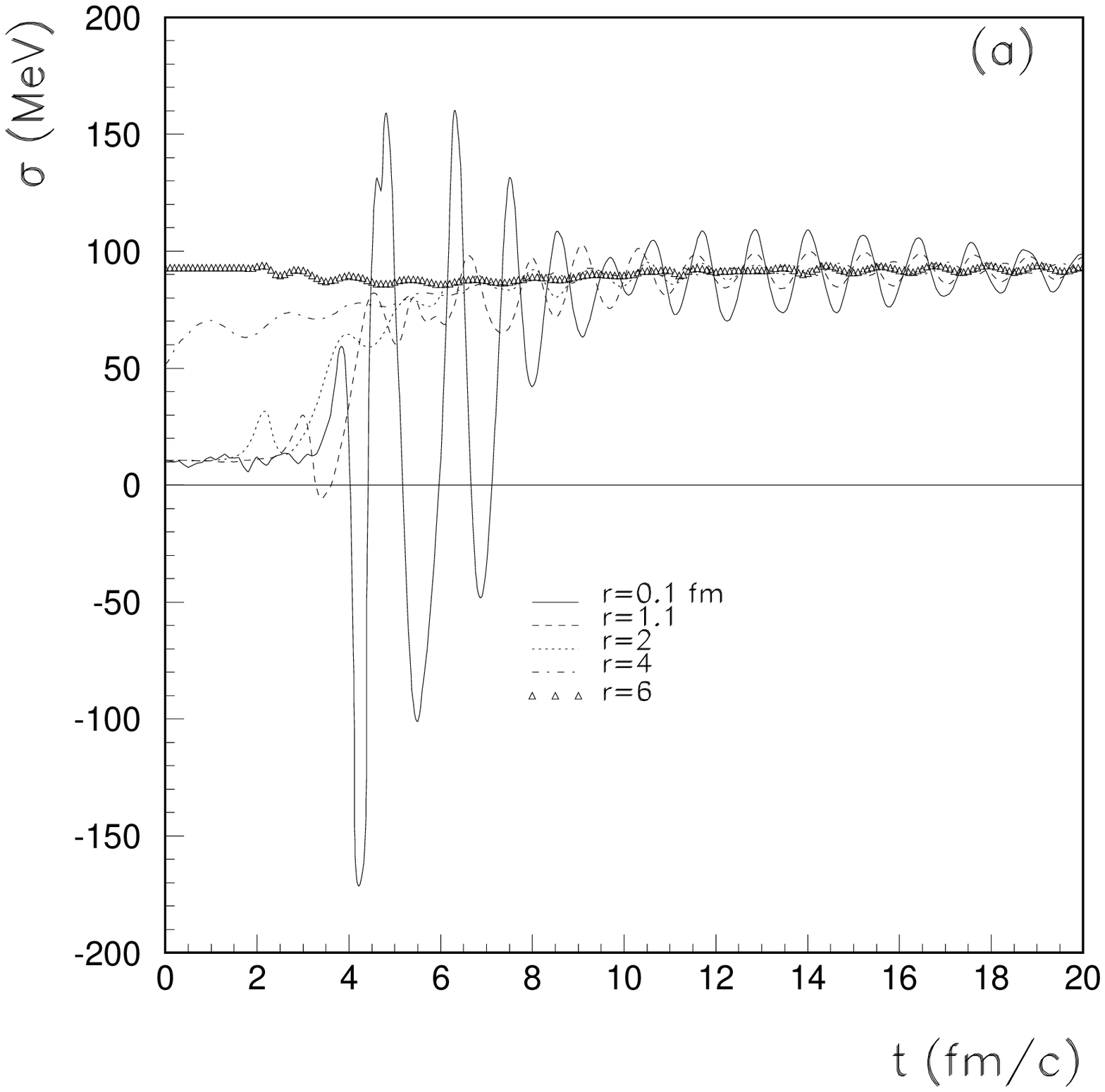}}
\end{figure}

\newpage
{\Large {\bf Abada/Birse, Fig.~2, part 2 of 2.} }
\vskip 1 cm
\begin{figure}[t]
\centerline{\psfig{figure=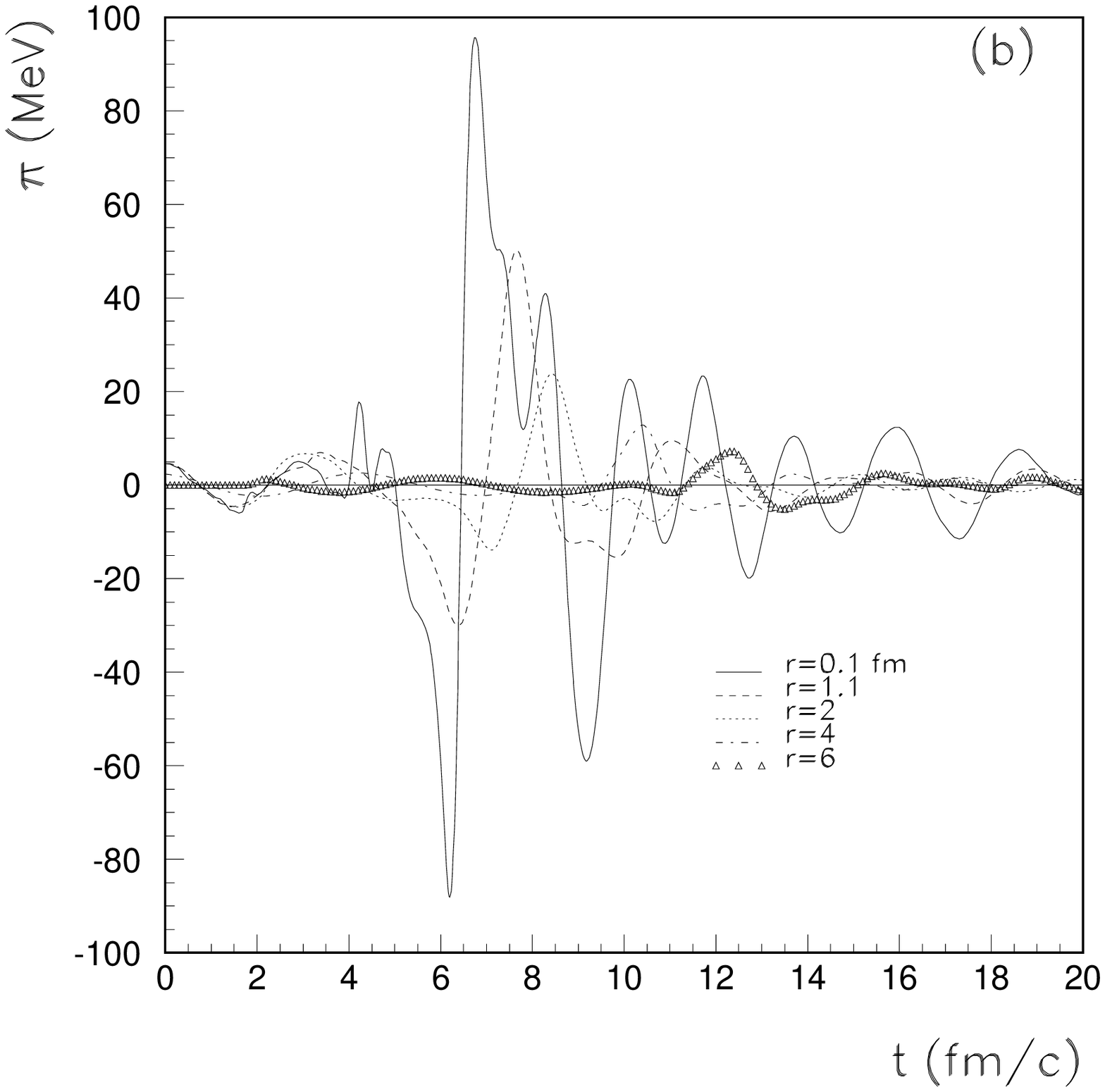}}
\end{figure}

\newpage
{\Large {\bf Abada/Birse, Fig.~3, part 1 of 4.} }
\vskip 1 cm
\begin{figure}[t]
\centerline{\psfig{figure=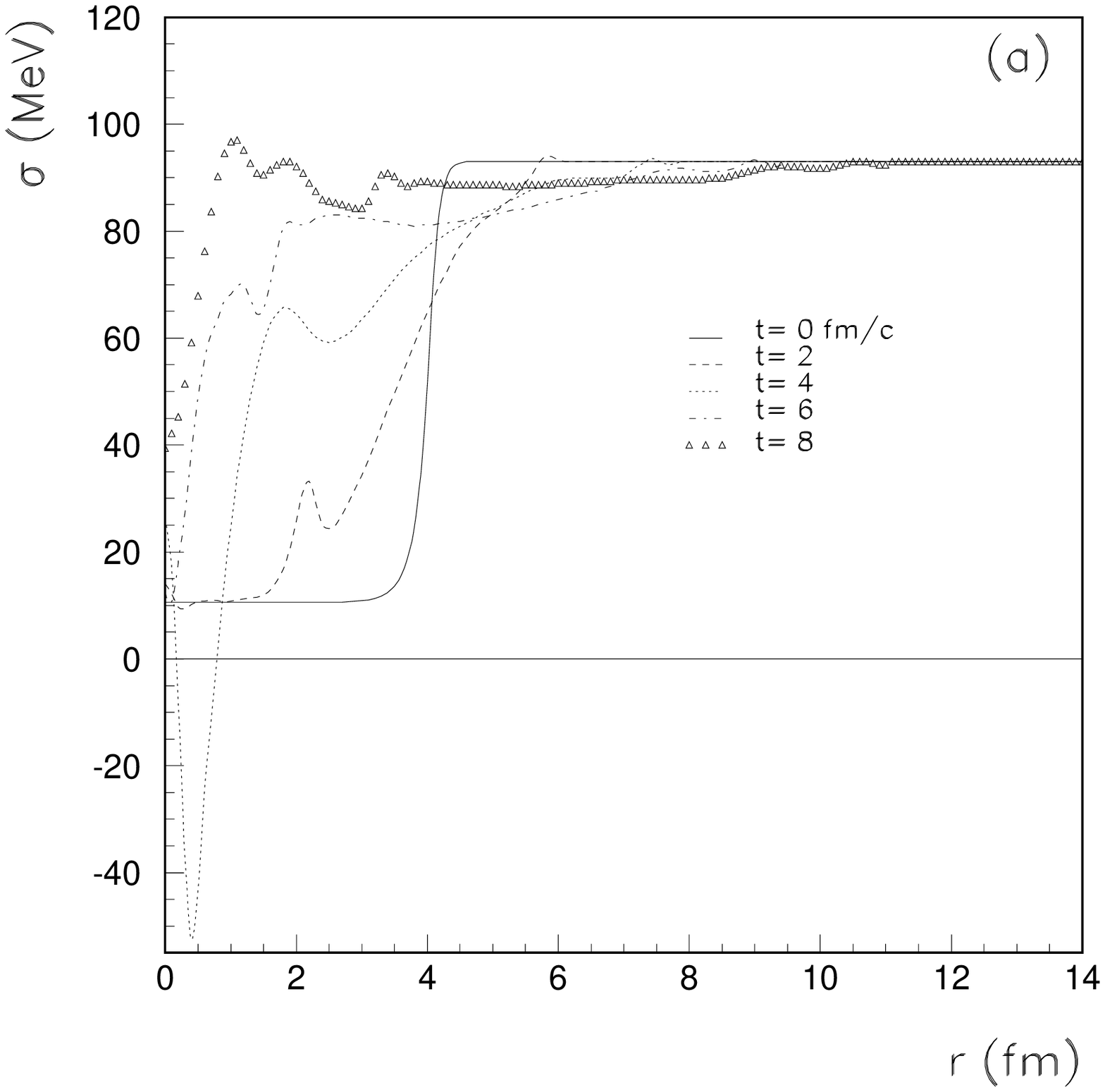}}
\end{figure}

\newpage
{\Large {\bf Abada/Birse, Fig.~3, part 2 of 4.} }
\vskip 1 cm
\begin{figure}[t]
\centerline{\psfig{figure=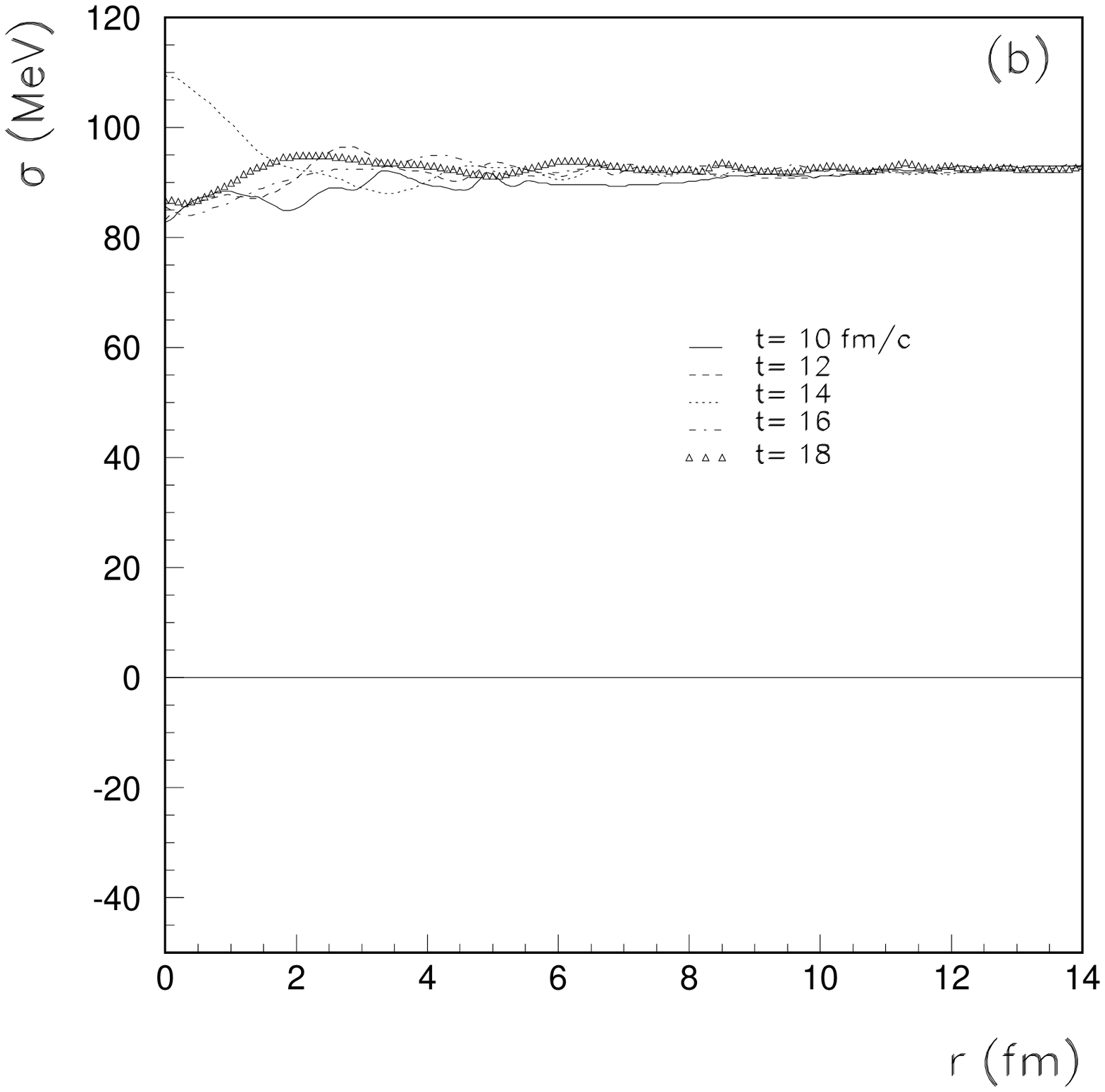}}
\end{figure}

\newpage
{\Large {\bf Abada/Birse, Fig.~3, part 3 of 4.} }
\vskip 1 cm
\begin{figure}[t]
\centerline{\psfig{figure=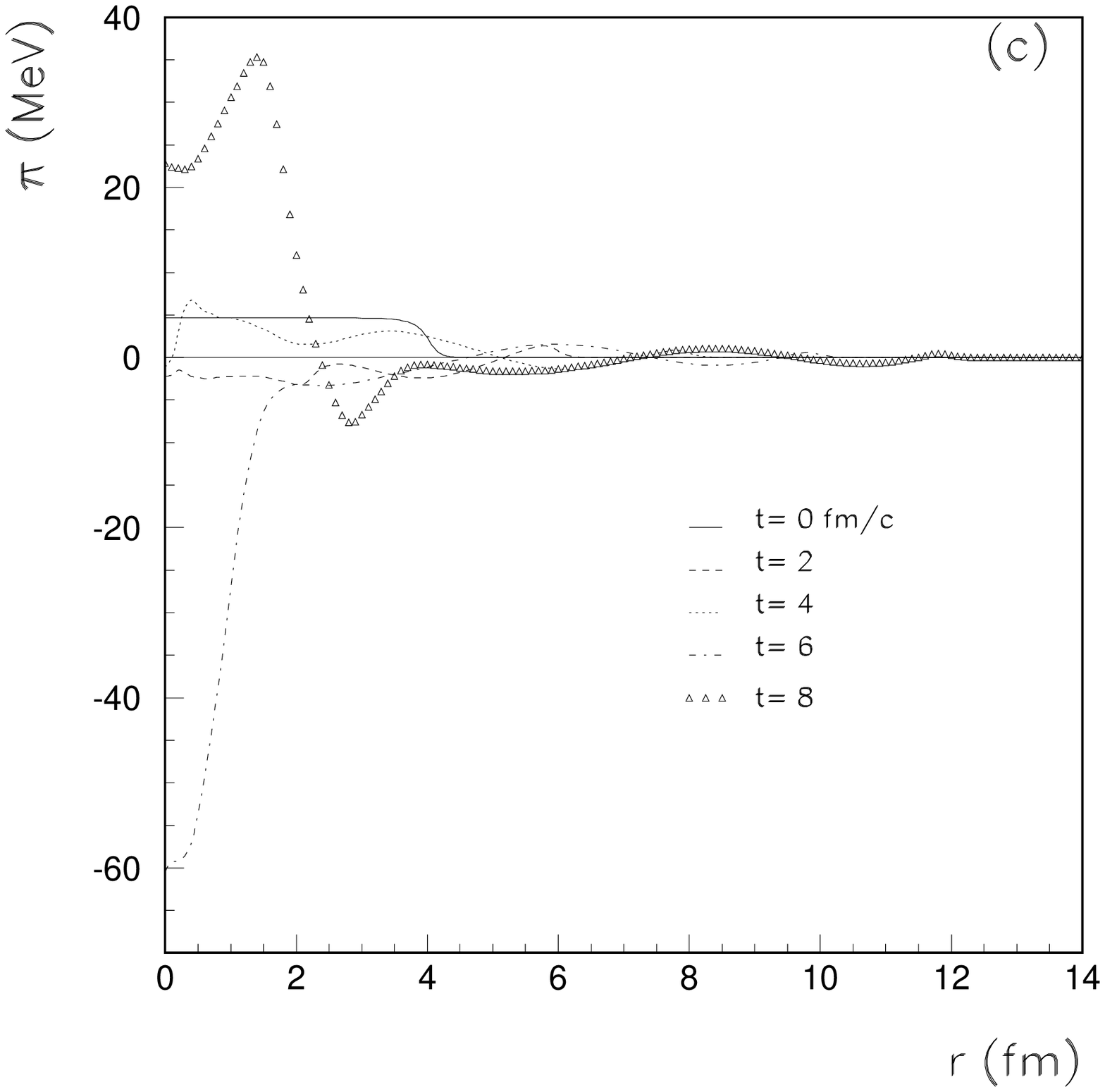}}
\end{figure}

\newpage
{\Large {\bf Abada/Birse, Fig.~3, part 4 of 4.} }
\vskip 1 cm
\begin{figure}[t]
\centerline{\psfig{figure=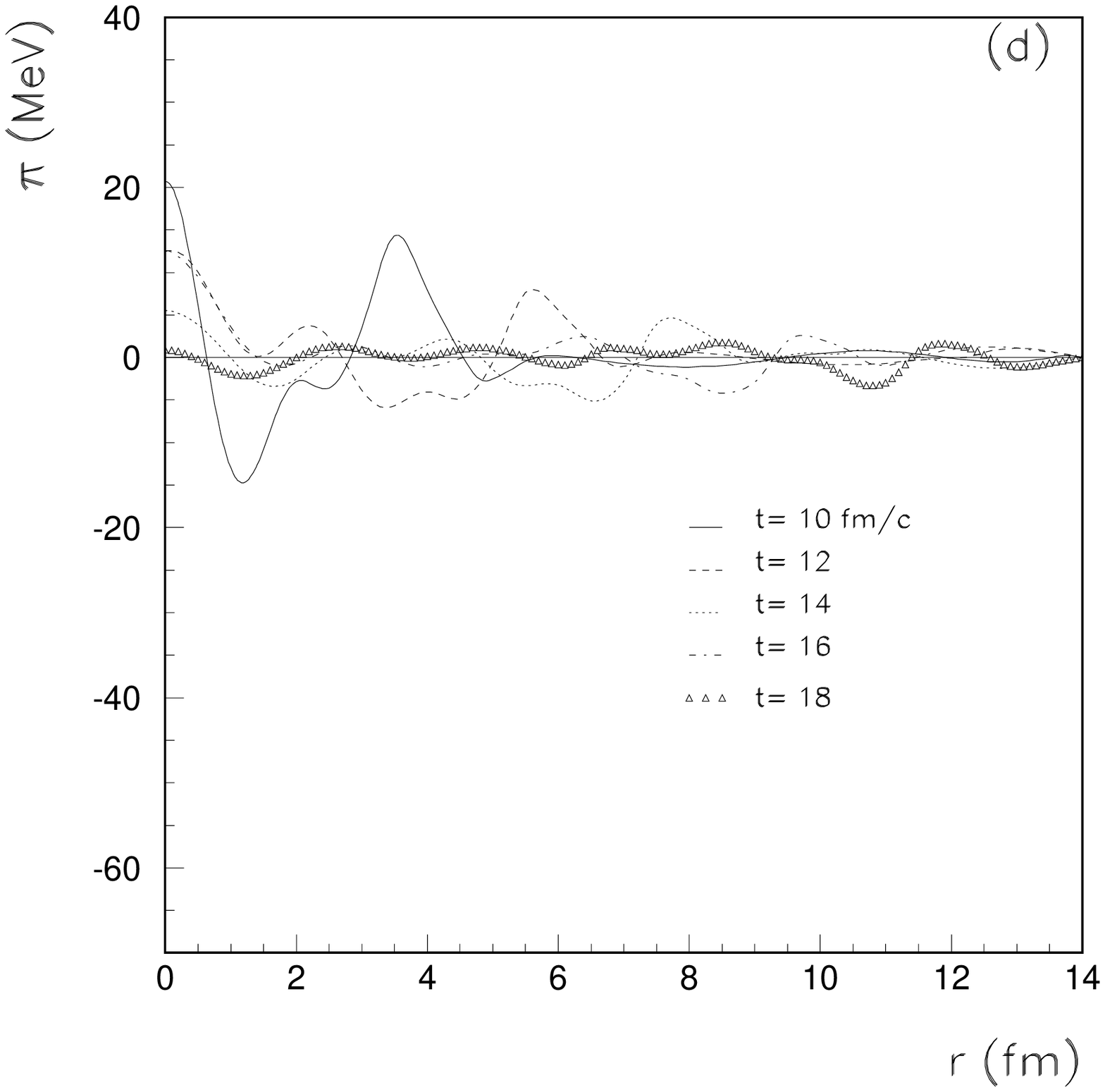}}
\end{figure}

\newpage
{\Large {\bf Abada/Birse, Fig.~4, part 1 of 2.} }
\vskip 1 cm
\begin{figure}[t]
\centerline{\psfig{figure=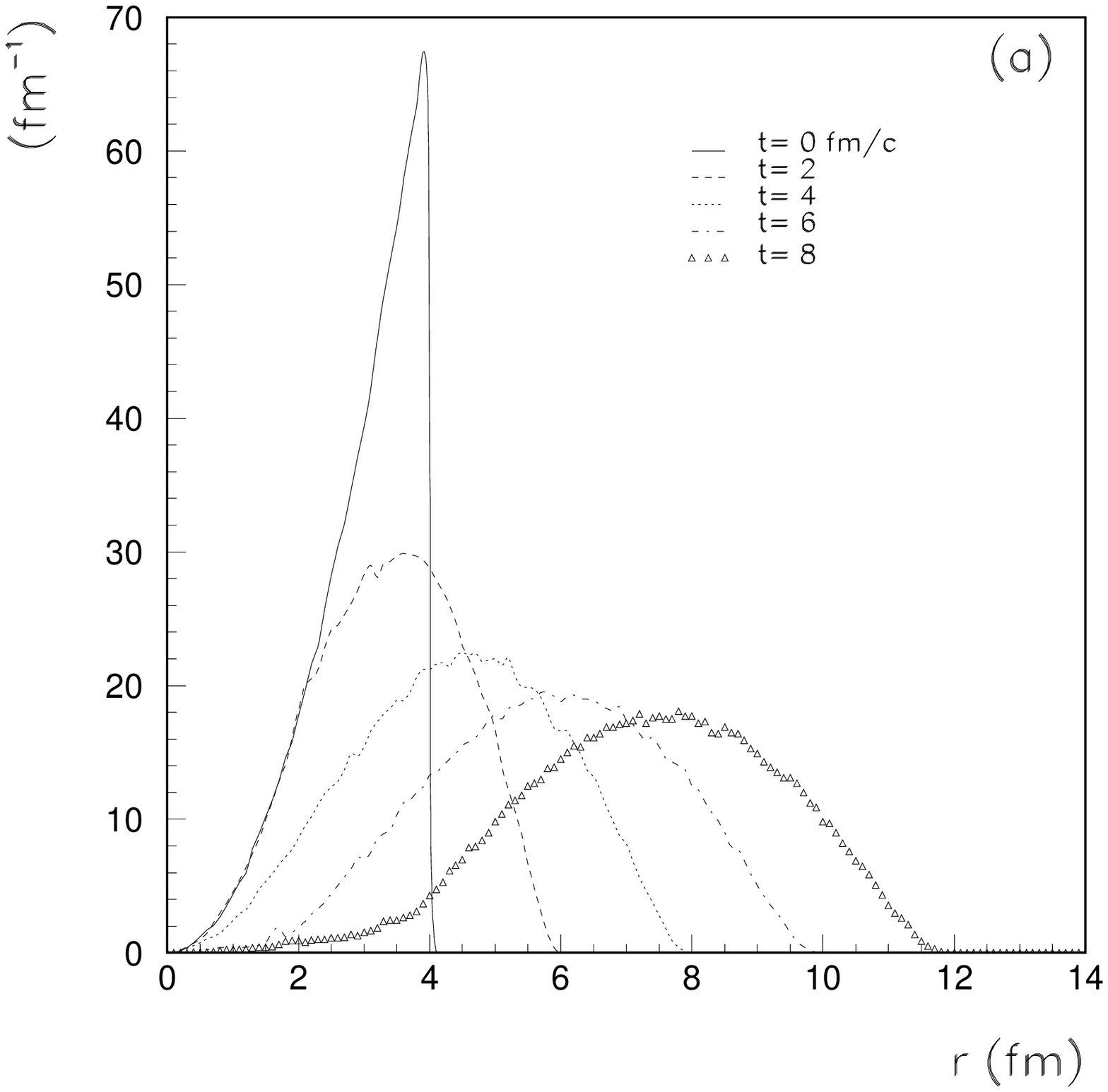}}
\end{figure}

\newpage
{\Large {\bf Abada/Birse, Fig.~4, part 2 of 2.} }
\vskip 1 cm
\begin{figure}[t]
\centerline{\psfig{figure=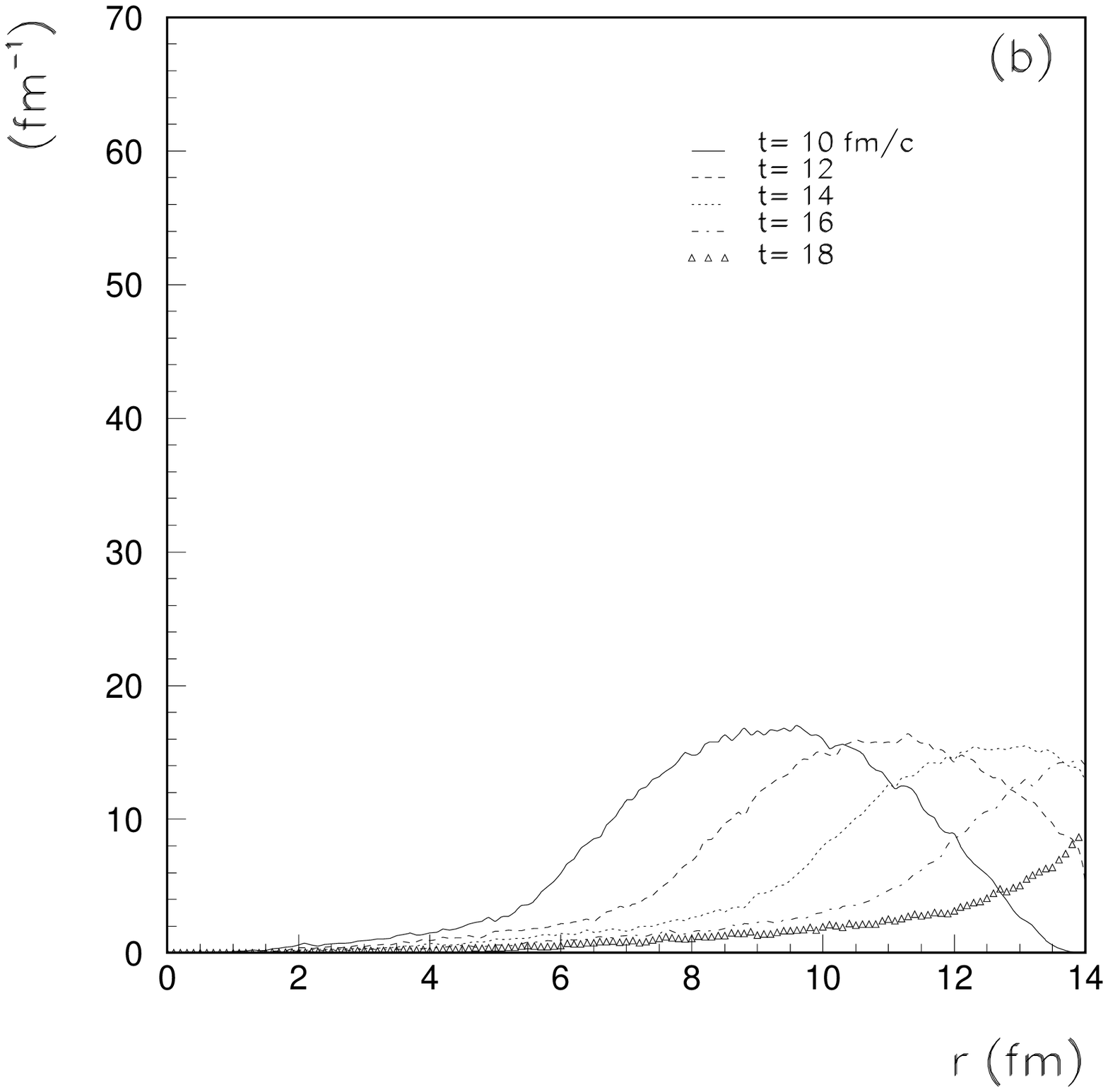}}
\end{figure}

\newpage
{\Large {\bf Abada/Birse, Fig.~5, part 1 of 2.} }
\vskip 1 cm
\begin{figure}[t]
\centerline{\psfig{figure=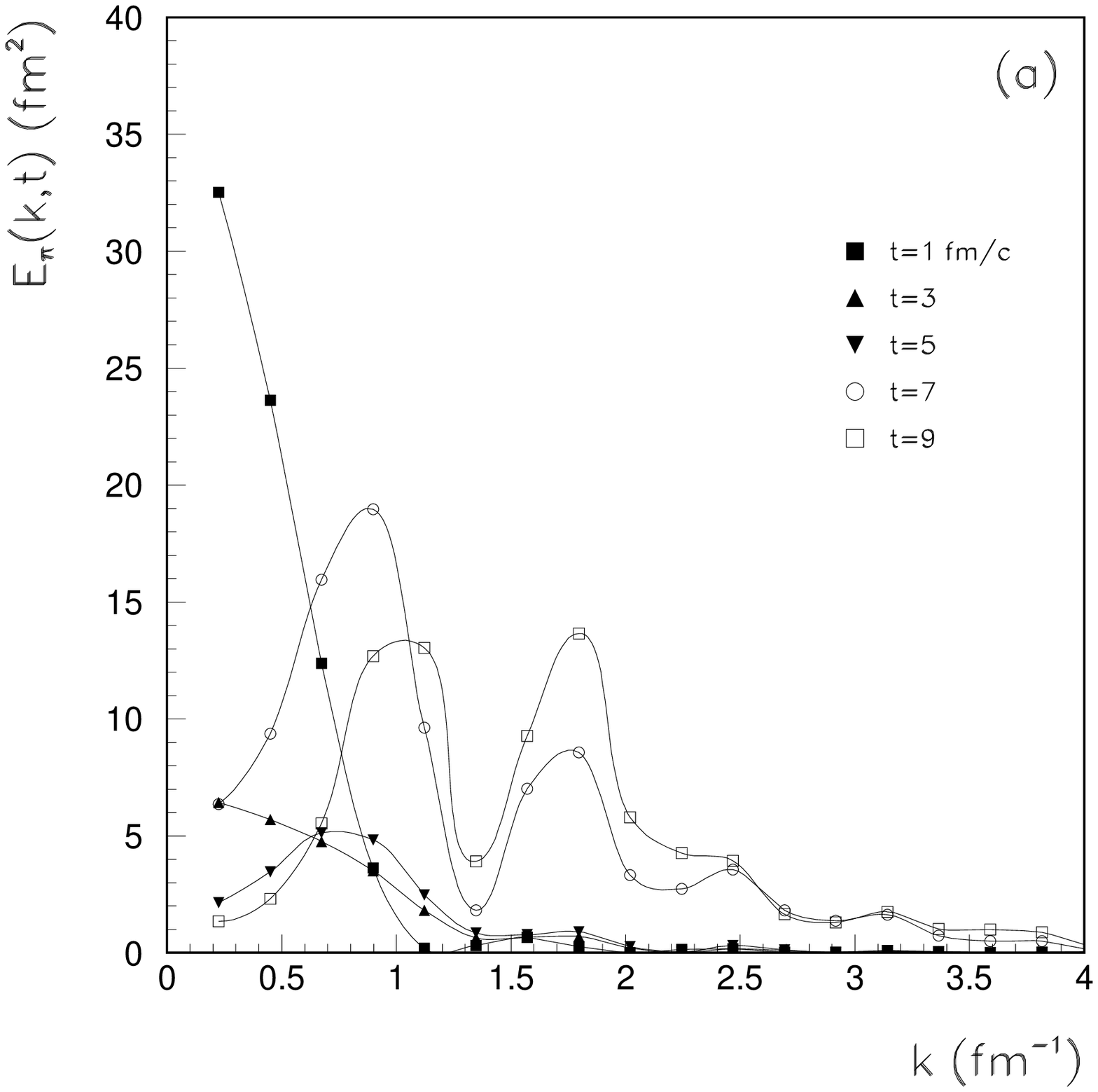}}
\end{figure}

\newpage
{\Large {\bf Abada/Birse, Fig.~5, part 2 of 2.} }
\vskip 1 cm
\begin{figure}[t]
\centerline{\psfig{figure=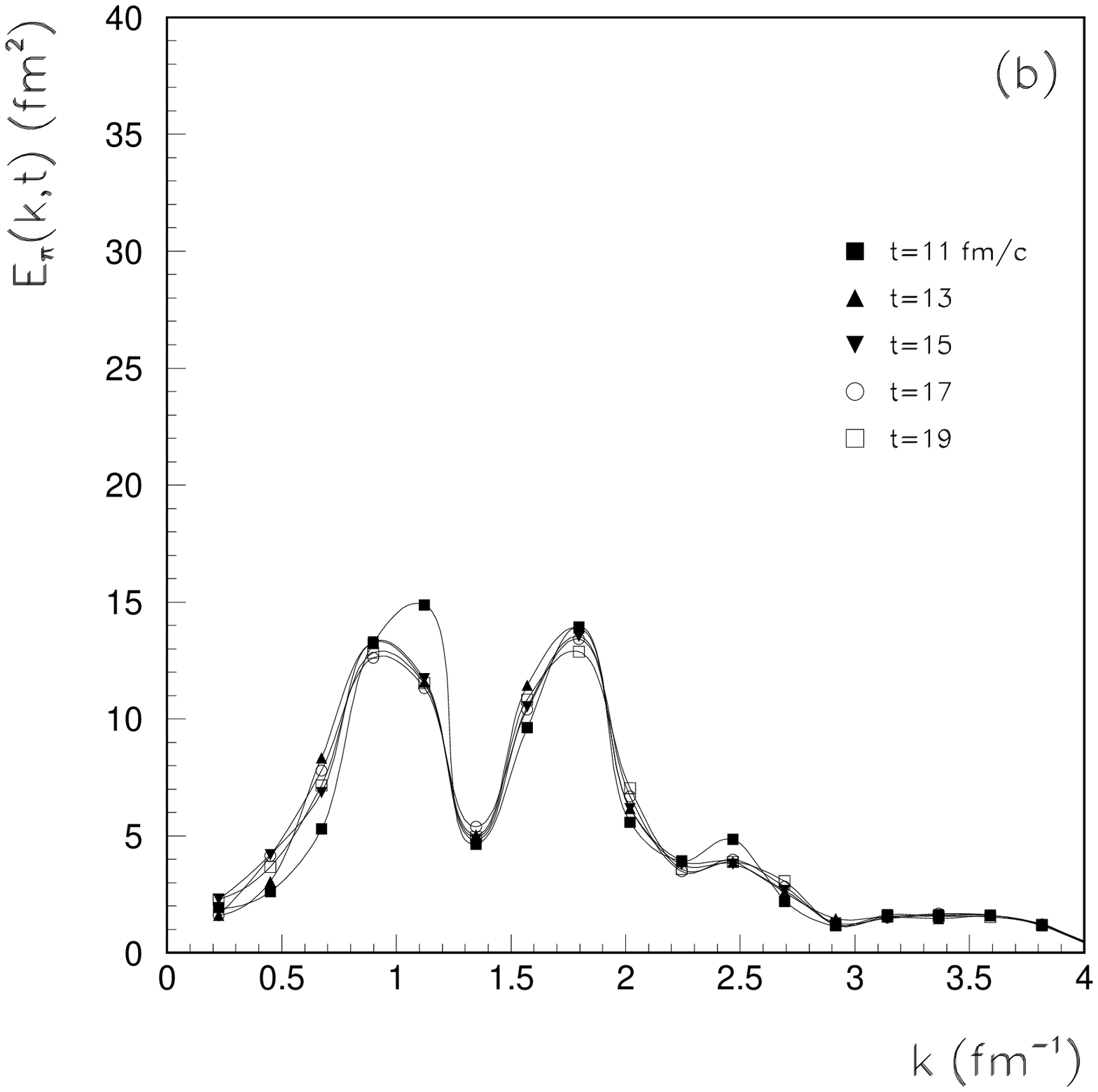}}
\end{figure}

\newpage
{\Large {\bf Abada/Birse, Fig.~6.} }
\vskip 1 cm
\begin{figure}[t]
\centerline{\psfig{figure=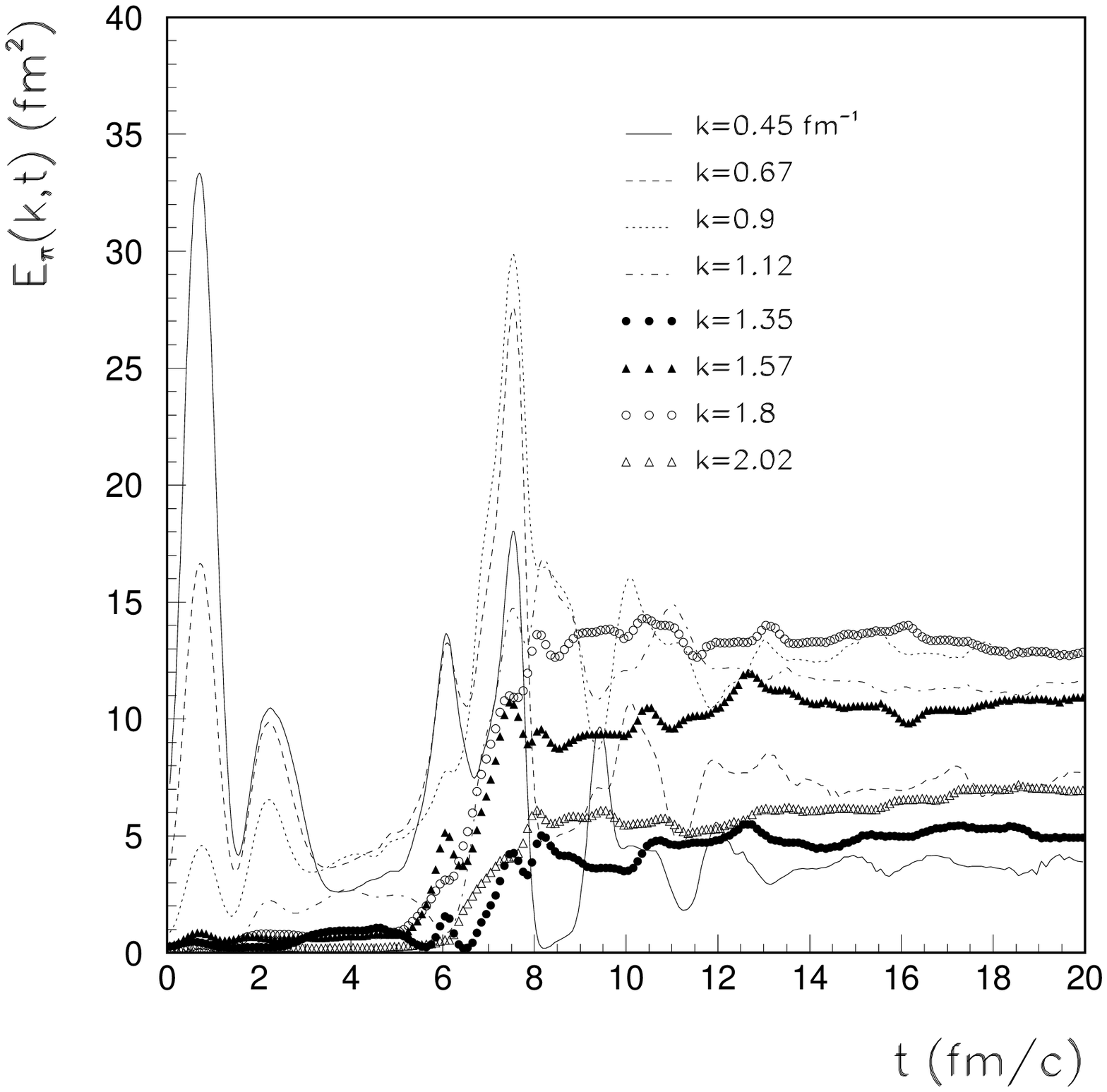}}
\end{figure}

\newpage
{\Large {\bf Abada/Birse, Fig.~7.} }
\vskip 1 cm
\begin{figure}[t]
\centerline{\psfig{figure=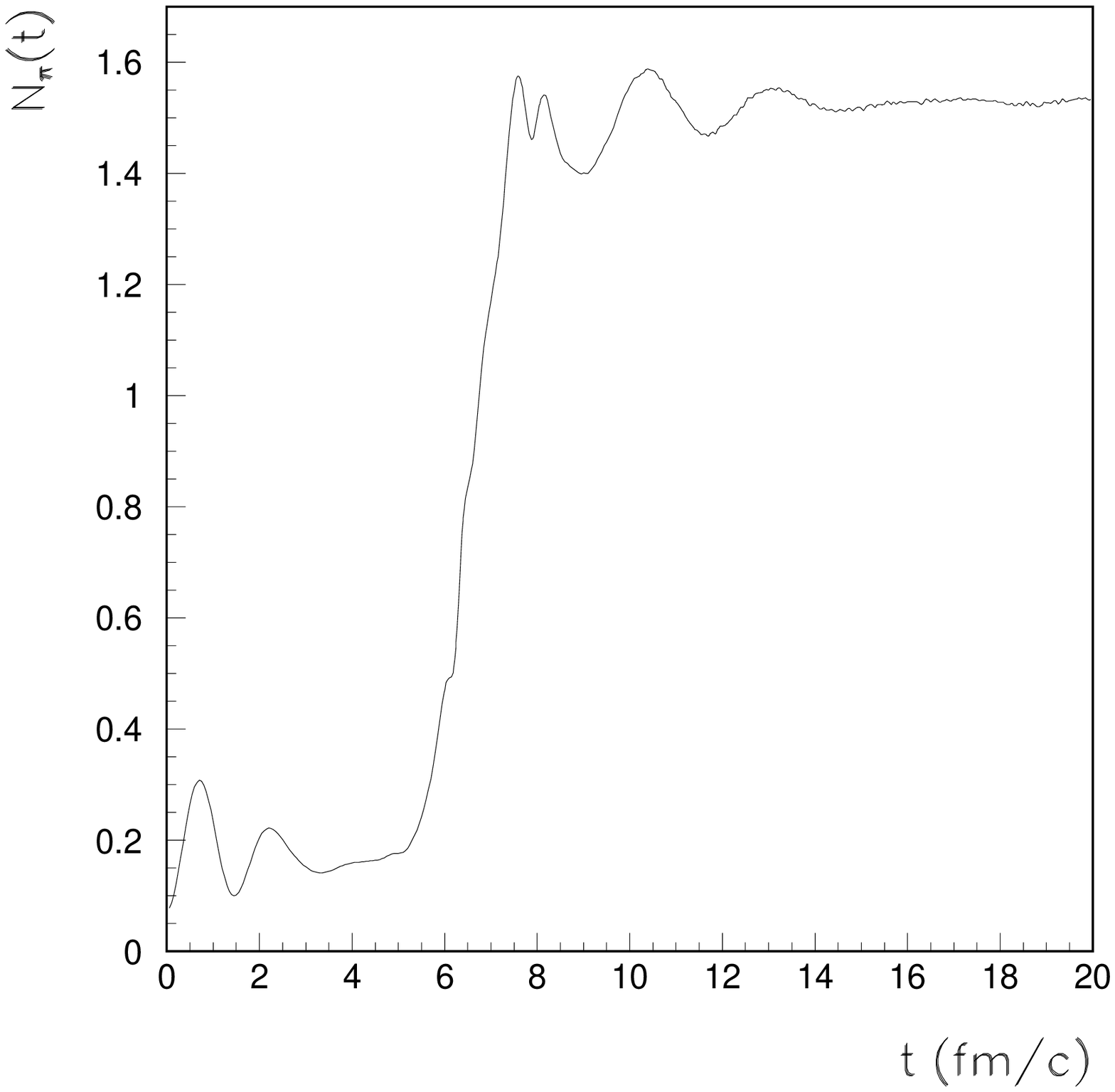}}
\end{figure}

\newpage
{\Large {\bf Abada/Birse, Fig.~8, part 1 of 2.} }
\vskip 1 cm
\begin{figure}[t]
\centerline{\psfig{figure=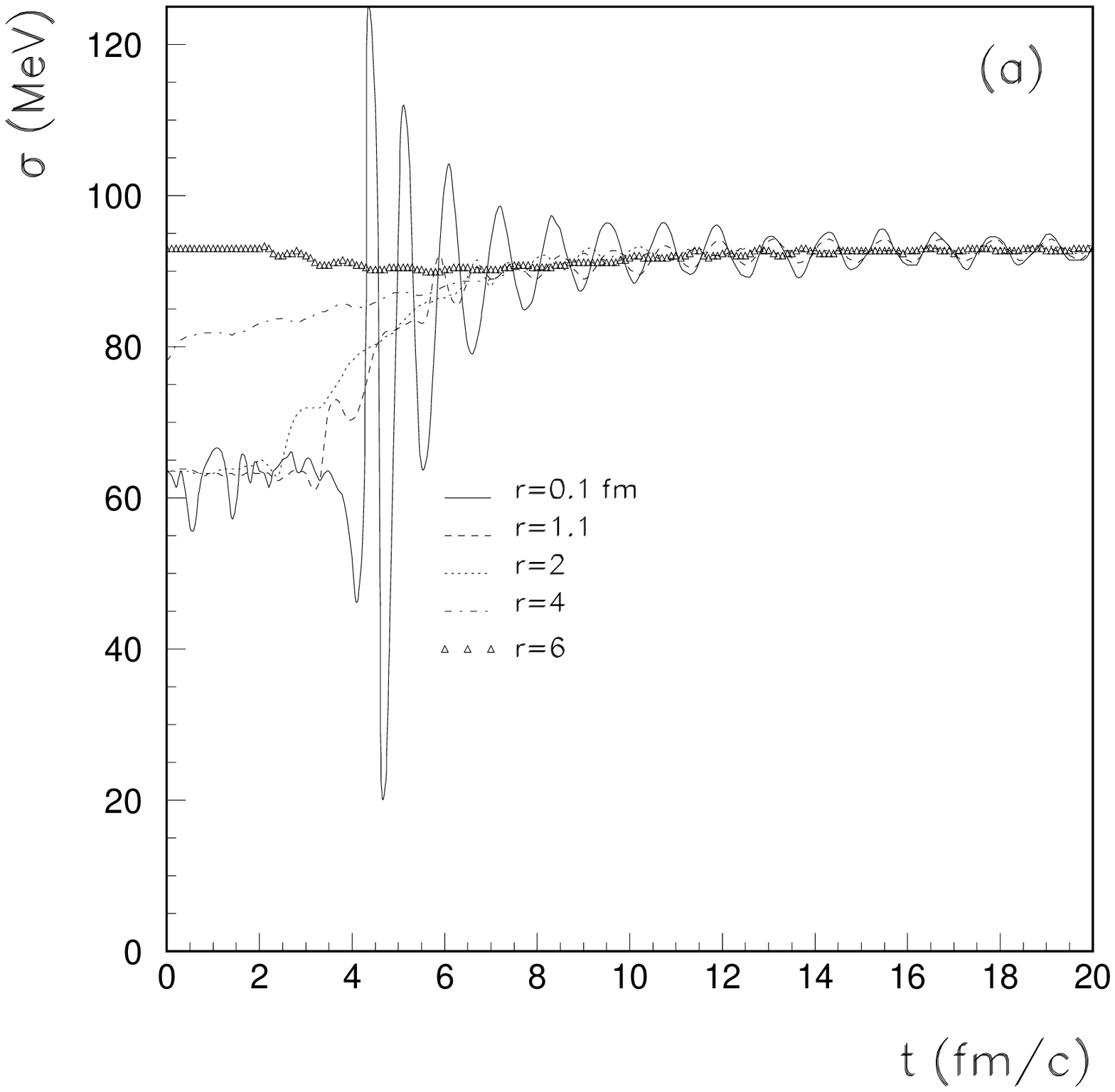}}
\end{figure}

\newpage
{\Large {\bf Abada/Birse, Fig.~8, part 2 of 2.} }
\vskip 1 cm
\begin{figure}[t]
\centerline{\psfig{figure=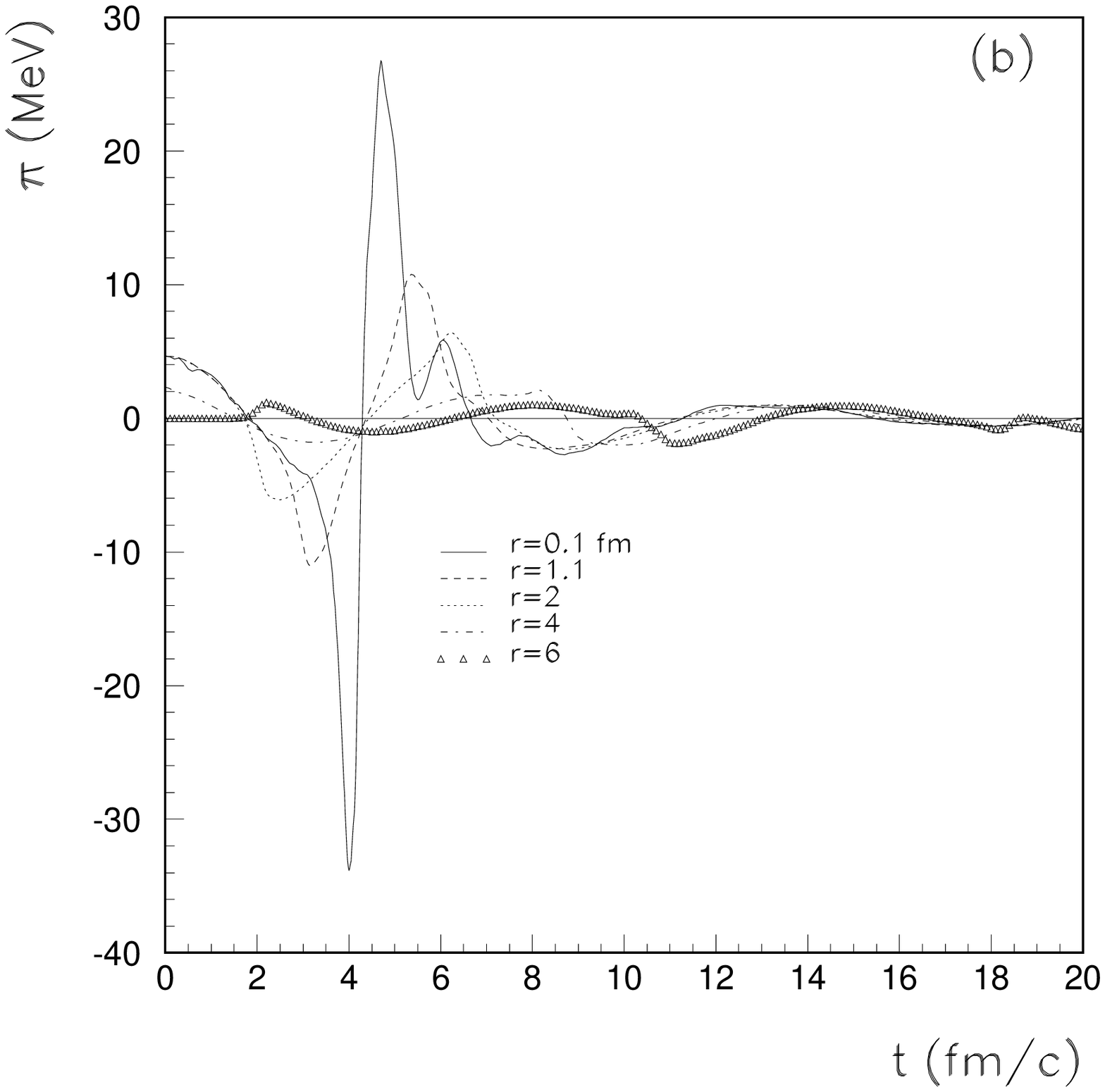}}
\end{figure}

\newpage
{\Large {\bf Abada/Birse, Fig.~9, part 1 of 2.} }
\vskip 1 cm
\begin{figure}[t]
\centerline{\psfig{figure=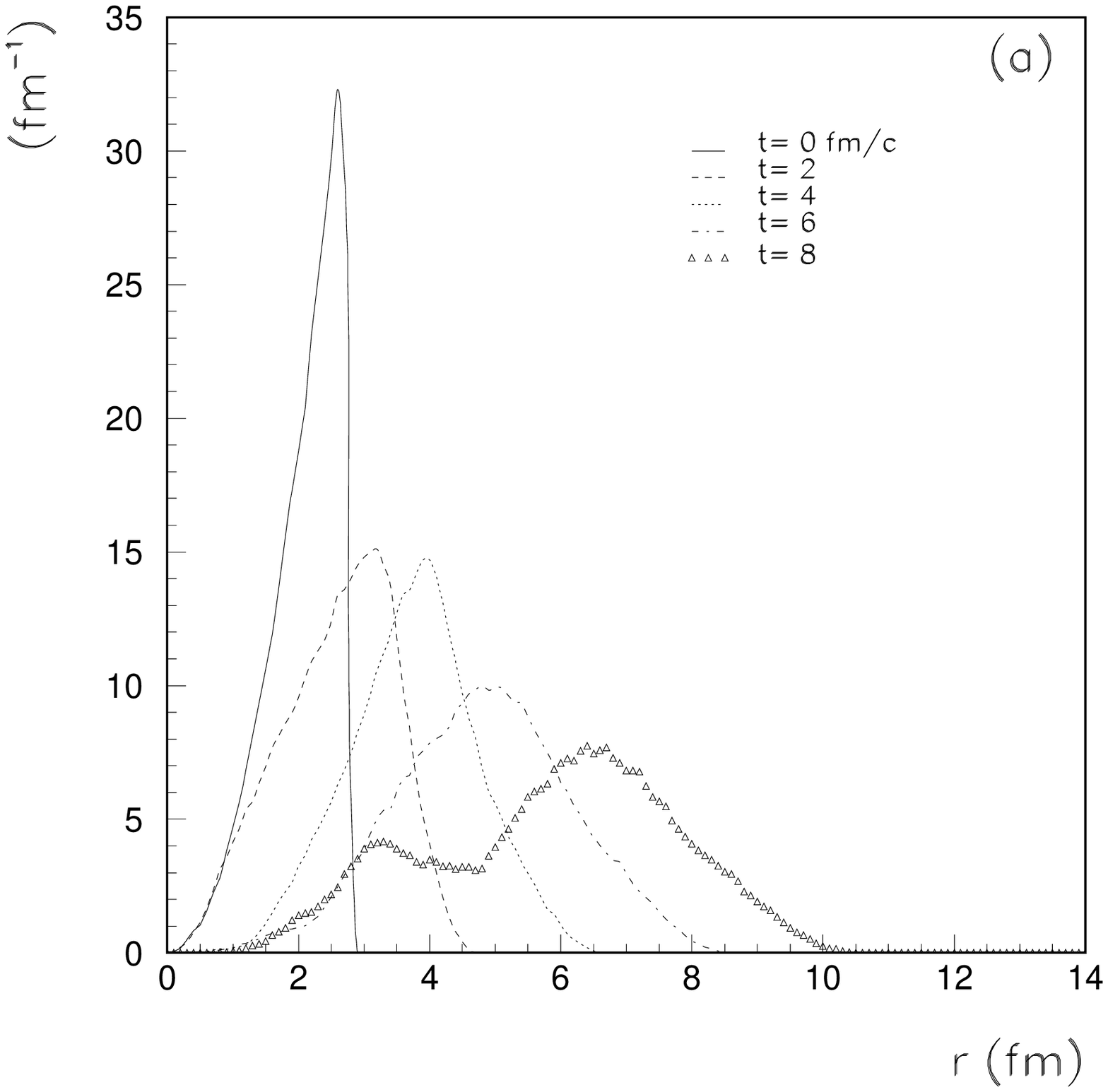}}
\end{figure}

\newpage
{\Large {\bf Abada/Birse, Fig.~9, part 2 of 2.} }
\vskip 1 cm
\begin{figure}[t]
\centerline{\psfig{figure=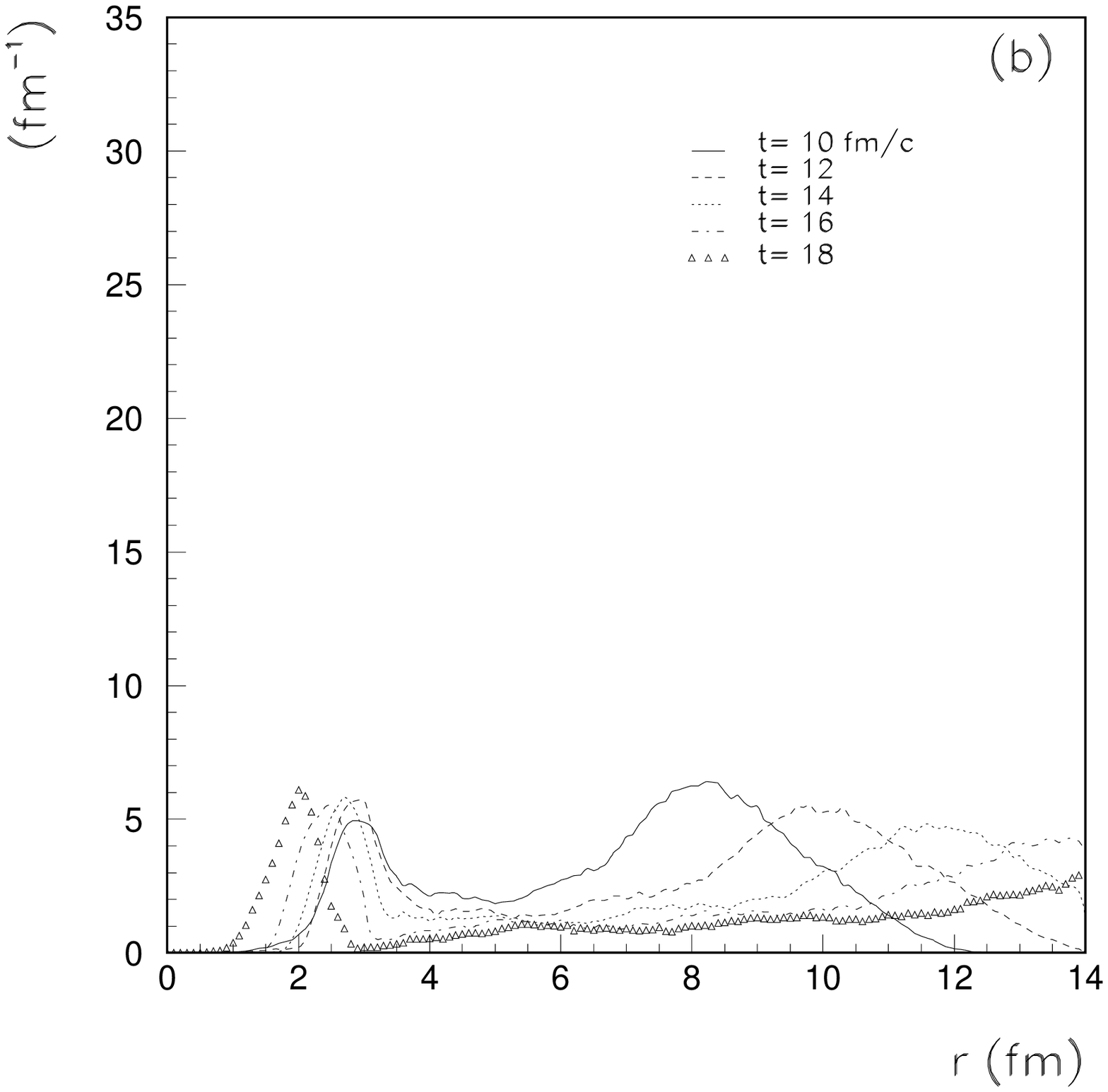}}
\end{figure}

\newpage
{\Large {\bf Abada/Birse, Fig.~10, part 1 of 2.} }
\vskip 1 cm
\begin{figure}[t]
\centerline{\psfig{figure=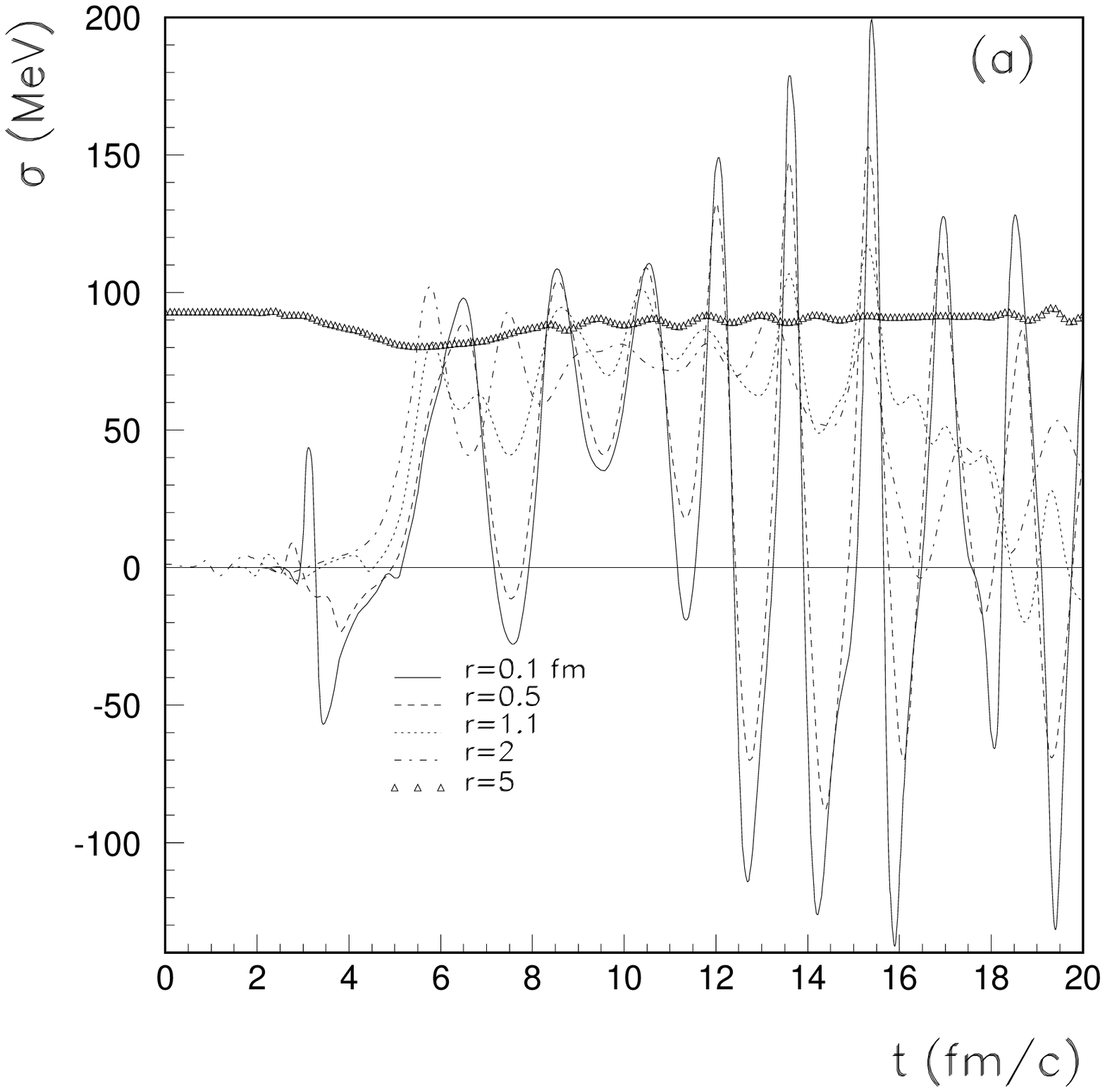}}
\end{figure}

\newpage
{\Large {\bf Abada/Birse, Fig.~10, part 2 of 2.} }
\vskip 1 cm
\begin{figure}[t]
\centerline{\psfig{figure=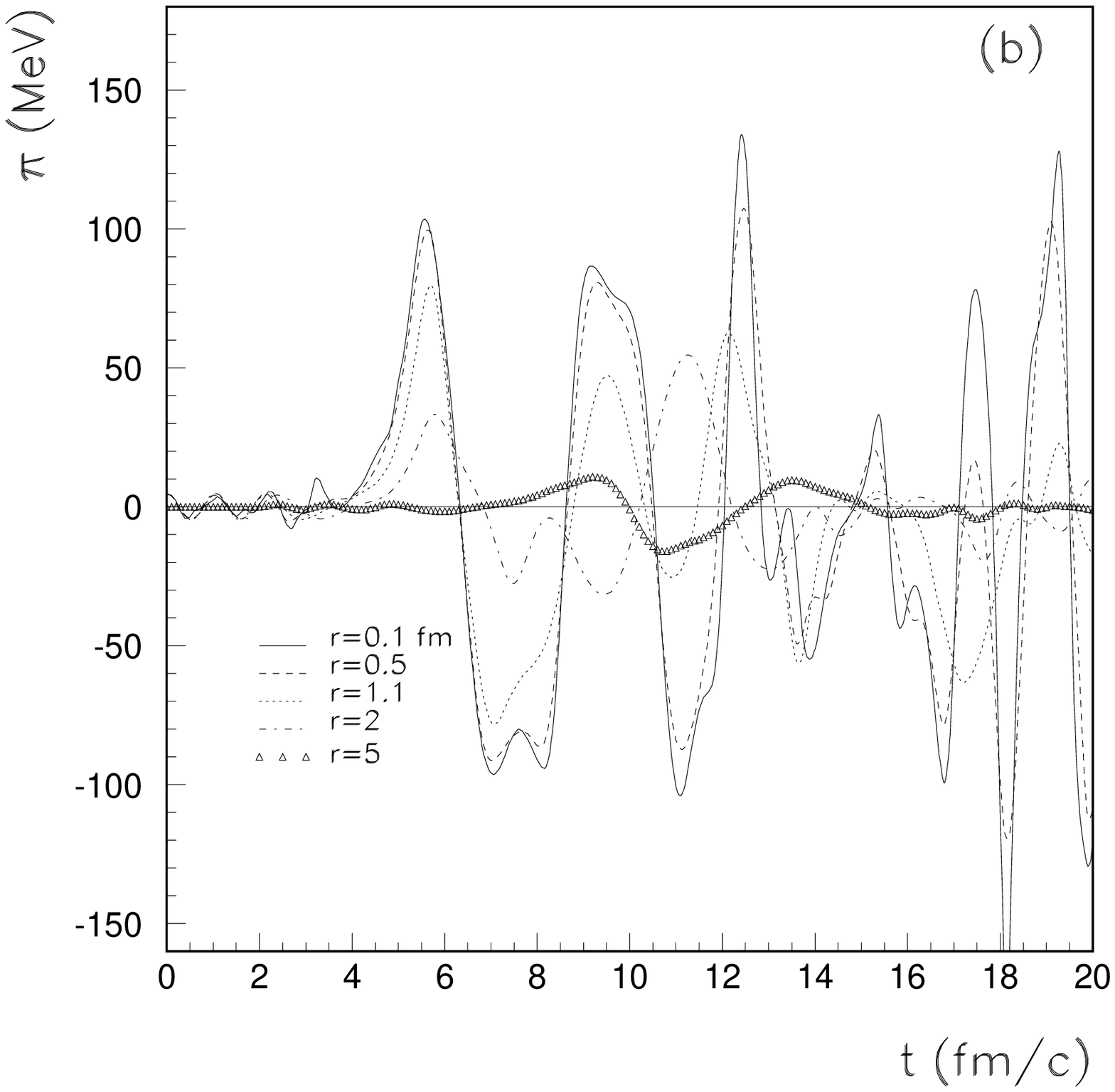}}
\end{figure}

\end{document}